\documentclass[fleqn,10pt]{wlscirep}
\title{The complete micromagnetic characterization of asymmetrically sandwiched ferromagnetic films}
\usepackage{color}

\author[1,2,*]{M. Kopte}
\author[2]{U. K. R\"o\ss ler}
\author[2]{R. Sch\"afer}
\author[1,2]{T. Kosub}
\author[1]{A. K\'{a}kay}
\author[1]{O. Volkov}
\author[1]{H. Fuchs}
\author[3]{E. Y. Vedmedenko}
\author[4]{F. Radu}
\author[2]{O. G. Schmidt}
\author[1]{J. Lindner}
\author[1]{J. Fa\ss bender}
\author[1,2,*]{D. Makarov}

\affil[1]{Helmholtz-Zentrum Dresden-Rossendorf e.V., Institute of Ion Beam Physics and Materials Research, Bautzner Landstrasse 400, 01328 Dresden, Germany}
\affil[2]{Leibniz Institute for Solid State and Materials Research Dresden e.V., Helmholtzstraße 20, 01069 Dresden, Germany}
\affil[3]{University of Hamburg, Institute of Applied Physics, Jungiusstrasse 11, 20355 Hamburg, Germany}
\affil[4]{Helmholtz-Zentrum für Materialien und Energie, Albert-Einstein-Str. 15, D-12489 Berlin, Germany}

\affil[*]{m.kopte@hzdr.de, d.makarov@hzdr.de}


\begin{document}

\begin{abstract}
The magnetic properties of ferromagnetic thin films down to the nanoscale are ruled by the exchange stiffness, anisotropies, and 
the effects of magnetic fields. As surfaces break inversion symmetry, an additional effective chiral exchange is omnipresent
in any magnetic nanostructure.
These so-called Dzyaloshinskii-Moriya interactions (DMIs) affect
all inhomogeneous magnetic state. These effects are mostly subtle, 
but can also be spectacular.
E.g., DMIs cause a chirality selection 
of the rotation sense and can fix the local rotation axis 
for the magnetization in domain walls, 
But, they can stabilize also entirely different
twisted magnetic structures. The chiral skyrmions 
a two-dimensional particle-like topological soliton 
is the ultimately smallest of these objects, which 
currently is targetted as a possible information
carrier in novel spintronic devices.
Observation and quantification 
of the chiral exchange effects provide for
the salient point in understanding magnetic properties in ultrathin films 
and other nanostructures. 
An easy and reliable method to determine the Dzyaloshinksii 
exchange constant as materials parameter of asymmetric 
thin films is the crucial problem.
Here, we put forth an experimental approach 
for the determination of the complete set of the 
micromagnetic parameters.
Quasi-static Kerr microscopy observations of domain wall creep motion and equilibrium sizes
of circular magnetic objects in combination with standard magnetometry are used to 
derive a consistent set of these materials parameters in polycrystalline ultrathin film 
systems, namely CrO$_x$/Co/Pt stacks.
The quantified micromagnetic model for these films identifies
the circular magnetic objects, as seen by the 
optical microscopy, as ordinary bubble domains with homochiral walls. 
From micromagnetic calculations, the chiral skyrmions stabilized by the DMI in
these films are shown to have diameters in the range 40 - 200~nm, too small
to be observed by optical microscopy.
\end{abstract}

\flushbottom
\maketitle

\thispagestyle{empty}


Chiral magnetic objects such as skyrmions\cite{Bogdanov1989, Bogdanov2001}, 
homochiral magnetic bubble domains\cite{Jiang2015} or homochiral 
domain walls (DW)\cite{Thiaville2012} promise to become new fundamental 
units for logic and memory devices.\cite{Kiselev2011, Parkin2008, Fert2013, Soumyanarayanan2016, Wiesendanger2016} 
The chirality of non-collinear magnetic inhomogeneities can be 
predetermined by the chiral exchange, an effect of the  relativistic spin-orbit  effect \cite{DZYALOSHINSKY1958,Moriya1960}. 
Ordinarily, this small effect only leads to a weak 
canting of moments in magnetically ordered materials. 
However, in certain acentric crystals the DMIs affect the magnetic order on larger 
length scale and stabilize chirally twisted states, as predicted theoretically by Dzyaloshinskii \cite{Dzyaloshinskii1964}, 
and later found in cubic helimagnets like MnSi \cite{BakJensen1980,NAKANISHI1980}.
The symmetry conditions for these chiral effects to occur are entirely 
general, and have been formulated by Dzyaloshinskii for ferromagnetic
and antiferromagnetic media.
As a surface or an interface unavoidably breaks inversion symmetry, 
these chiral effects are always present 
in any magnetic nanostructure.\cite{Fert1990,Crepieux1998,Bogdanov2001,Roessler2006}
Recognition of the importance and ubiquity of these effects 
has been slow, and experimental investigations of 
chiral magnetic behavior in thin magnetic films have started only 
in recent years. Early observations demonstrated 
chiral magnetic spiral ground-states 
in a most simple antiferromagnetic monolayer film \cite{Bode2007},
or the selection of on handedness during the formation of
vortex states in magnetic nanodisks \cite{Im2012}.\\
Current interest in the chiral exchange effects in nanostructures 
is centered on their peculiar ability to stabilize localized solitonic states.
The chiral DMIs have been theoretically shown to stabilize two-dimensional 
radial solitons \cite{Bogdanov1989,Bogdanov1994,Bogdanov2001}.
These solitons are double-twisted magnetizati configurations with a unique
sense of rotation that are now called chiral magnetic skyrmions. 
These multidimensional solitons are fundamentally different from ordinary 
circular domains, which are the well-known magnetic bubbles \cite{Kiselev2011}.
%
%
The swirl-like configuration of the chiral skyrmion is described by a radial profile where neighboring spins are noncollinearly twisted, both in radial and azimuthal direction. Homochiral bubble domains consist of a homogeneously magnetized circular area confined by a homochiral DW. Although both objects have the same topological charge, a unit skyrmion number, they display different dynamical properties, stability, and susceptibility to pinning\cite{Fert2013}. The performance of skyrmion-based devices, e.g. storage density and the operation speed will be determined by the physical nature of the specific type of the chiral object, which in turn imposes stringent requirements on the functional magnetic layer. For instance, in films with isolated free skyrmions the modulated states, like one-dimensional spiral phases and skyrmion lattice states, must be suppressed by strong uniaxial anisotropies\cite{Butenko2010}, while a homochiral DMI is preserved.
Design of thin magnetic film systems carrying isolated chiral magnetic skyrmions with defined properties
requires not only breaking of inversion symmetry, but control of the strength of the chiral 
DM exchange and of the magnetic anisotropies.
\begin{figure*}[!b]
\centering
\includegraphics[width=.8\linewidth]{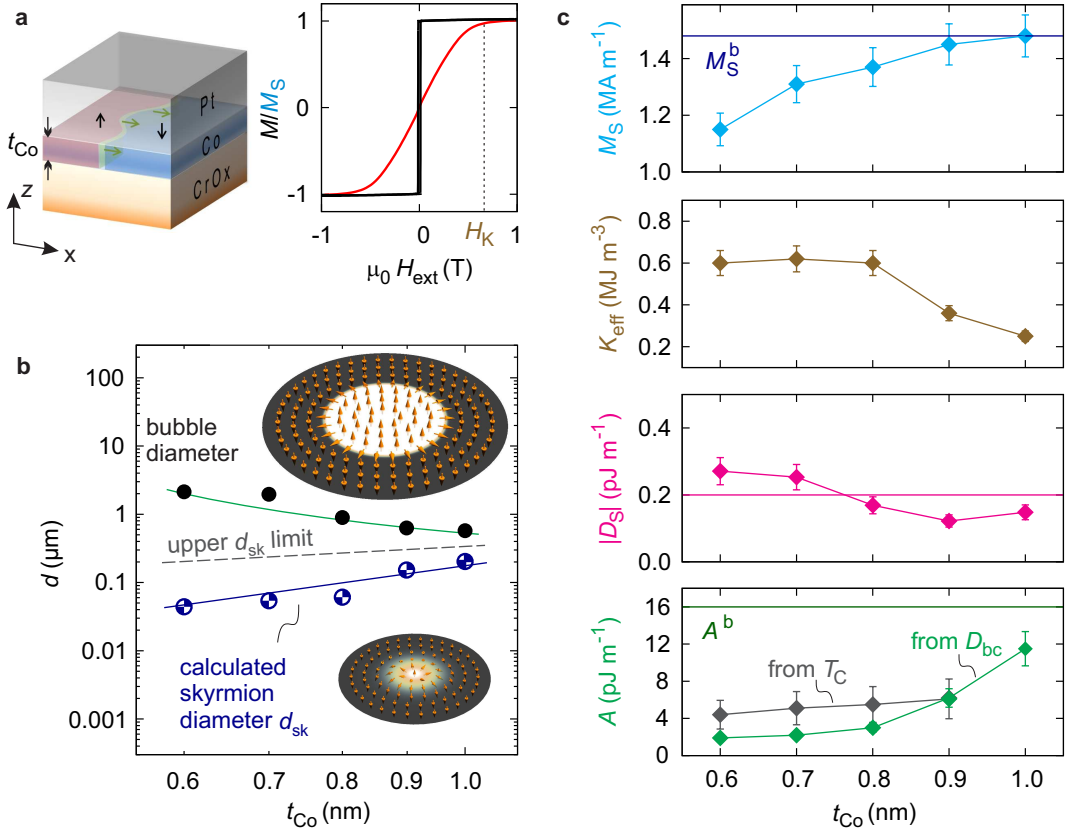}
\caption{Full micromagnetic characterization of //CrOx/Co($t_\textrm{Co}$)/Pt trilayers with varying cobalt layer thickness $t_\textrm{Co}$. 
\textbf{(a)} Sketch of the trilayer stack and the easy ($z$) and hard ($x$) axis hysteresis loops of the magnetization obtained by anomalous Hall magnetometry normalized to the saturation magnetization $M_\textrm{S}$. \textbf{(b)} The diameter scaling the magnetic objects allows to distinguish between skyrmions and homochiral bubbles domains. The calculated skyrmion diameter $d_\textrm{sk}$ 
from the measured set of micromagnetic parameters. The upper limit of $d_\textrm{sk}$ is obtained from simulations with the bulk exchange value. \textbf{(c)} Significant cobalt layer thickness ($t_\textrm{Co}$) dependence of the complete set of micromagnetic parameters including $M_\textrm{S}$, magnetic anisotropy $K_\textrm{eff}$, surface induced Dzyaloshinskii-Moriya interaction $D_\textrm{S}$ and exchange parameter $A$. The sizable deviation from the bulk value $A^b$ is supported by the approximate estimation of $A$ from the Curie temperatures $T_\textrm{C}$ (Supplementary).}
\label{Fig:results}
\end{figure*}
Promising technological material platforms supporting 
individual localized chiral objects even at room temperature\cite{Jiang2015, Boulle2016, Woo2016} are multilayer system generally comprised of an ultra-thin ferromagnetic (FM) layer sandwiched between two heavy metal (HM) layers\cite{Je2013,Hrabec2014,Lavrijsen2015,Di2015,Woo2016} or one HM and a metal oxide (MO) layer.\cite{Belmeguenai2015, Cho2015, Kim2015a, Vanatka2015, Boulle2016} 
Recently, in search of chiral skyrmions in such film systems, the creation 
and observation of circular magnetic objects have been reported 
displaying a very wide range of typical diameters, e.g., from around 50~nm\cite{Legrand2017} up to 2~$\mu$m\cite{Jiang2015}, although  experimentally determined Dzyaloshinskii constant are almost the same.
The skyrmion diameter is approximately fixed by the chiral modulation length $L = A/D$, 
given by the ratio of direct exchange $A$ to Dzyaloshinskii constant $D$ for the chiral DMIs.  
The uniaxial anisotropy $K_u$ and demagnetization effect modifies the profiles and size of 
the skyrmions to a certain extent \cite{Bogdanov1994}. 
Together with the saturation magnetization $M_s$, 
this complete set of micromagnetic materials parameters 
in a thin ferromagnetic film also determines 
the properties of the magnetic domain walls
and the magnon dynamics of the collinear ferrromagnetic state. 
As sufficiently anisotropic magnetic films have a collinear
ground-state, the effects of chiral DMIs are hidden
and can be traced only indirectly by observing their
effects on non-collinear magnetization configurations 
and their behavior.
Various approaches have been proposed, how to determine 
the Dzyaloshinskii constant.\cite{Butenko2009,Cortes-Ortuno2013, Langer2016}
By now a standard method is to measure the asymmetry of 
magnon dispersion in thin films. However this method 
requires rather high quality of the films, which is 
not always practiable for complex multilayered and polycrystalline films.
Observations of quasi-static inhomogeneities affected by 
the DMIs is an alternative.
Microscopic observations of wall profiles, vortex core sizes, or skyrmions 
as excitations are difficult, as extremely high resolution must
be achieved to measure tiny differences, e.g., in a left and right
handed magnetic vortex or wall section \cite{Butenko2009}.
Also, it is experimentally not straighforward to stabilize
DWs with different handedness in a film material.

Therefore, the method currently employed relies on 
observations of DW mobility,
as the interface energy of the wall is modified by the combination
of the chiral DMI and appropriate external oblique fields \cite{Je2013,Hrabec2014}.
This method only requires magneto-optical Kerr microscopy
with the resolution of a light microscope.
This method is an experimentally appealing 
and can provide a routine approach to 
determine the chiral magnetic properties of candidate 
platforms for studies on isolated magnetic
skyrmions. 
However, it is highly indirect 
and relies on a precise determination 
of all other micromagnetic materials parameters.
At present, there is no agreement in reported data regarding 
the strength of the DMIs and the size of skyrmions in 
asymetrically sandwiched thin films, Dzyaloshinskii
constants determined by different methods are not in 
agreement, and there is no consensus about various 
influences of different surfaces and cappings or substrate
layers on the DMIs in asymmetric sandwiches 
\cite{Belmeguenai2015, Cho2015, Kim2015a,Vanatka2015, Di2015a,Boulle2016}.\\
Here, we employ the quasi-static approach by observing DW creep to determine the DMI strength. By using a modified analysis method, we show how it can be used  to quantify the full set of the micromagnetic parameters. We develop a measurement routine, which in conjunction with the proper theoretical framework, allows to obtain the Dzyaloshinskii constnats from a sequence of Kerr microscopy measurements without any ambiguous assumption on the exchange constant. When the DMI strength is determined, the exchange parameter can be deduced from characteristic diameters of circular magnetic objects. 
Localized homochiral objects can be identified either as skyrmions or magnetic bubbles. This can be achieved by following the scaling behavior of the object size with the ferromagnetic layer thickness.\\ 
This approach is validated on a //CrOx/Co/Pt trilayer system combining the features of (i) stacking-order inversion and (ii) a novel metal-oxide joining the family of so far investigated systems based on aluminum, magnesium and gadolinium oxide.\cite{Belmeguenai2015, Cho2015, Kim2015a,Vanatka2015, Di2015a,Boulle2016} In this system important technology-relevant criteria like all-electric accessibility for read-out purposes and easy film preparation are ensured from an inverted growth of the film stack (substrate/MO/FM/HM). This implies a reduced optimization effort on MO layer preparation and avoids chemical interactions at the Co/MO interface during deposition. Furthermore, CrOx is introduced as an alternative MO system, in order to clarify the existence of an enhancing contribution of MO interfaces to the DMI strength.\cite{Boulle2016, Belabbes2016} The cobalt layer thickness is varied to assess the scaling properties of the DMI and homochiral magnetic objects.\\
From the characteristic scaling dependence of the object diameter with respect to the cobalt layer thickness, we infer that the investigated systems support homochiral bubbles, rather than skyrmions. The DMI strength in our samples has comparably small values, $D_\textrm{S}=-(0.20\pm 0.09)\,\textrm{pJ\,m}^{-1}$ which can yet support homochiral magnetic textures. Remarkably, the obtained exchange stiffness in the studied sample series are significantly diminished compared to the bulk value of cobalt. We further find an unambiguous evidence that interface quality or more generally microstructural features and defects modify the effective DMI mechanism.  On the other hand, no empirical evidence is found for a contribution of the Co/MO interface to the DMI strength. 

\section*{Results}

\begin{figure*}[!b]
\centering
\includegraphics[width=.8\linewidth]{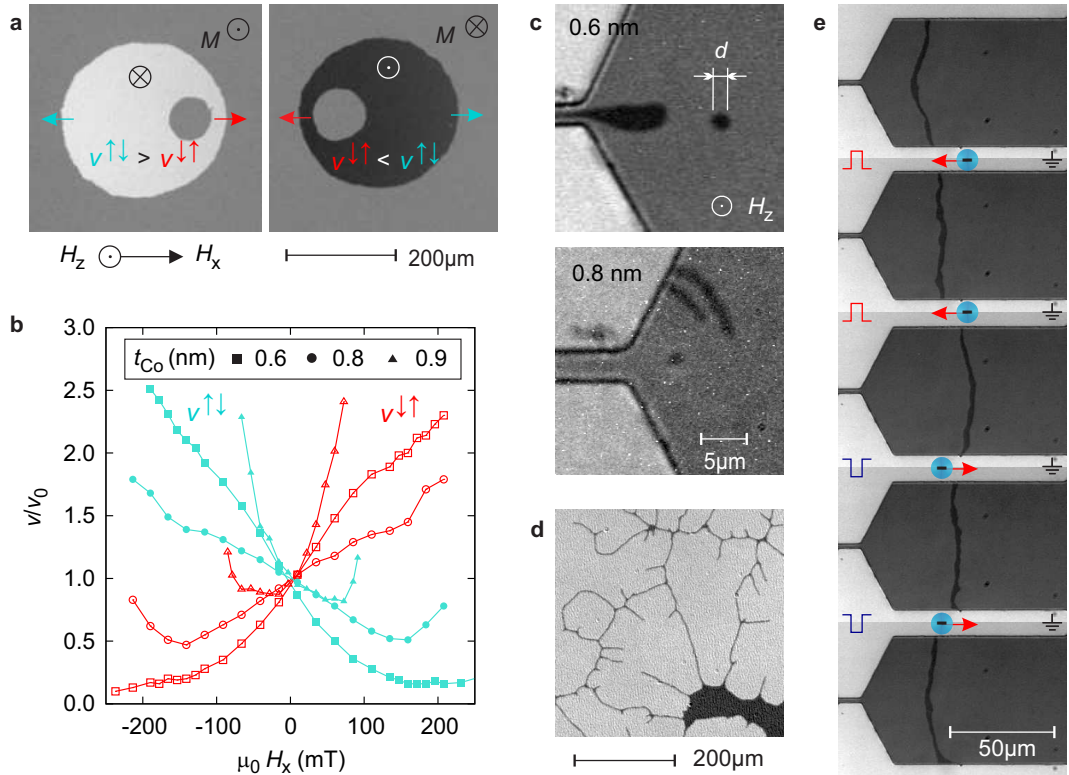}
\caption{Quasi-static Kerr microscopy measurements as tool to determine the exchange parameters of asymmetrically sandwiched ferromagnetic thin films. \textbf{(a)} Field driven domain growth in presence of a hard axis field $H_\textrm{x}$ reveal asymmetries of the creep velocities for $\uparrow\downarrow$ and $\downarrow\uparrow$ DWs as function $H_\textrm{x}$ \textbf{(b)}. As the minimum of $v(H_\textrm{x})$ is deficient measure of the DMI field, an $A$-independent DMI quantification approach is developed essentially requiring one asymmetry ratio of the creep velocities $v^{\uparrow\downarrow}/v^{\downarrow\uparrow}$ in the low region of $H_\textrm{x}$. 
\textbf{(c)} Circular magnetic objects generated in the scheme of Jiang et al.\cite{Jiang2015} According to the characteristic scaling (Fig.~\ref{Fig:results}b)  the  objects classify as homochiral bubbles. The observed collapse diameter $d$ is used to deduce the exchange constant $A$ from bubble domain theory. The homochiral nature of the DWs induced by the DMI can be inferred from the observation of \textbf{(d)} winding pairs formed during field driven DW creep motion and \textbf{(e)} a coherent current-induced DW motion. Both $\uparrow\downarrow$ and $\downarrow\uparrow$ DWs move against the electron motion. A sequence of current pulses of different polarity and a current density of $10^{6}\,\textrm{A\,cm}^{-2}$ was applied for 40~ms.}
\label{Fig:Kerr}
\end{figure*}

In the following the full sequence of measurement and analysis steps for the determination of the micromagnetic materials parameters is described.\\
The DMI quantification method used here relies on the analysis of the field-driven evolution of domain patterns. The remagnetization process in the studied samples is generally governed by the nucleation of domains and subsequent propagation of DWs through the entire film. The DW velocities $v$ obey the creep law $v= v_0\,\exp[\zeta_0/(\mu_0H_\textrm{z})^{1/4}]$ with the scaling parameters $v_0$ and $\zeta_0$ (Supplementary~\ref{sec:creep}). With an additionally applied field ($H_\textrm{x}$) perpendicular to the easy axis, the domain growth exhibits an asymmetric distortion as shown in figure~\ref{Fig:Kerr}a. This is a consequence of the combined modification of the DW energy by $H_\textrm{x}$ and the DMI field $H_\textrm{DM}=D_\textrm{S}/(\mu_0\,M_\textrm{S}\,\sqrt{A/K_\textrm{eff}}\,t_\textrm{Co})$.\cite{Thiaville2012} The DMI field is a function of all micromagnetic parameters including saturation magnetization $M_\textrm{S}$, anisotropy energy $K$, exchange parameter $A$ and DMI strength $D_\textrm{S}$. 
As shown in Fig.~\ref{Fig:Kerr}a the asymmetry of the domain growth is most pronounced along the axis of $H_\textrm{x}$. By evaluating the dependence of the DW creep velocities $v$ of $\uparrow\downarrow$ and $\downarrow\uparrow$ DWs a $v(H_\textrm{x})$-curve is recorded as plotted for different cobalt layer thicknesses in  figure~\ref{Fig:Kerr}b. In an simple approximation \cite{Je2013, Lavrijsen2015,Vanatka2015} and only for a limited number of cases $H_\textrm{DM}$ can be directly read  from a usually broad minimum of the $v(H_\textrm{x})$-curve. To rescale the DMI field to the Dzyaloshinskii constant for the surface-induced DMIs $D_\text{S}$, a precise knowledge of all intrinsic parameters, especially the exchange constant $A$, is required. Since $A$ is difficult to determine, usually the bulk value of $A^b$=16~pJ/m for cobalt is assumed. Furthermore, ambiguities have been raised, if the method is performed in the creep regime at high in-plane fields ($H_\textrm{x}$).\cite{Vanatka2015}\\
Therefore, we first derive a revised analysis that is able to determine $D_\text{S}$ independently from the value of the exchange stiffness $A$, based on the original idea \cite{Je2013,Hrabec2014} for the quantification of the DMI strength (see methods section). By focusing the evaluation of the DW velocity asymmetry on the low $H_\textrm{x}$-field region the aforementioned ambiguities can be avoided. This allows to establish relation between the value of the Dzyaloshinskii constant and the ratio of asymmetric DW velocities $v^{\uparrow\downarrow}/v^{\downarrow\uparrow}$ which are measured in dependence on the obliquely applied field.
The relation reads
\begin{equation}
D_\text{S}  = \frac{64\,\ln(2)}{\pi^2}\,\ln\left(\frac{v^+(H_\text{x},H_\text{z})}{v^-(H_\text{x},H_\text{z})}\right)\,\frac{(\mu_0\,H_\text{z})^{1/4}}{\zeta_0\,H_\text{x}} \,M_\text{S}\,K_\text{eff}, 
\label{eq:DS}
\end{equation} as derivied in Methods. 
Here, only the asymmetry ratio of the DW-velocities $v^{\uparrow\downarrow}/v^{\downarrow\uparrow}$ for one field combination of $H_\textrm{x}$ and $H_\textrm{z}$ and the creep parameter $\zeta_0$ are essential inputs from experiment. The two intrinsic magnetic parameters $M_\textrm{S}$ and $K_\textrm{eff}$ can be obtained from elementary magnetometry.
Hence, a generally applicable approach is provided, even if no clear minimum can be obtained.\\

%
%
The typical size of the homochiral magnetic objects (Fig~\ref{Fig:Kerr}c) allows for the determination of one unknown parameter of the micromagnetic parameter set, here the exchange parameter $A$. Depending on the scaling properties of the object diameter (see Fig~\ref{Fig:results}b and Supplementary) the analysis has to be carried out either in the framework of isolated magnetic bubble domains\cite{Hubert1998} or skyrmion theory\cite{Leonov2016}.\\
In the first case of magnetic bubble domains the diameter is expected to follow an inversely proportional trend to the cobalt thickness. In a correctly chosen experimental regime with fields just below the collapse field the generated objects eventually vanish at the size of the  characteristic collapse diameter $d_\textrm{bc}=d/t_\textrm{Co}$. According to bubble domain theory\cite{Hubert1998} the stability of the bubble is related to the ratio of the DW energy $\sigma_{DW}$ and the shape anisotropy $K_\textrm{D} = \mu_0\,M_\textrm{S}^2/2$. We point out, that in the presence of DMI the modified DW energy of $4\sqrt{A\,K_\textrm{eff}} - \pi\,D_\textrm{S}/t_\textrm{Co}$ has to be considered and therefore the $A$-independent determination of $D_\textrm{S}$ is an inevitable premise. After the calculation of the stability criterion $S_\textrm{bc}(d_\textrm{bc})$ for the measured collapse diameter (see methods section) the exchange parameter $A$ remains as the only free parameter, if $M_\textrm{S}$ and $K_\textrm{eff}$ are known, and can be calculated by the following relation
\begin{equation}
A= \frac{t_\text{Co}^2}{16\,K_\text{eff}} \left[\mu_0\,M_\text{S}^2\,S_\text{bc}(d_\text{bc})+\pi\,D_\text{S}\right]^2.
\label{eq:Aquant}
\end{equation}
Only if a scaling of the diameter $\propto (D_\textrm{S}\,t_\textrm{Co})^{-1}$ is observed, the objects classify as skyrmions. Micromagnetic simulations according to Leonov et al.\cite{Leonov2016} with known $M_\textrm{S}$ and $K_\textrm{eff}$ can be performed to determine the exchange parameter by only varying $A$ until the resulting diameter matches the experimentally observed skyrmion size.
\begin{table*}[t]
\centering
\caption{Dependence of magnetic parameters of //CrOx/Co($t_\textrm{Co}$)/Pt samples on the cobalt layer thickness $t_\textrm{Co}$  including saturation magnetization $M_\textrm{S}$, anisotropy field $H_\textrm{K}$, effective anisotropy constant $K_\textrm{eff}$, creep parameter $\zeta_0$, coercive field $H_\textrm{c}$, Curie temperature $T_\textrm{C}$, exchange parameter $A$, domain wall width $\Delta$, the cobalt thickness normalized DMI constants $D_\textrm{S}$ and the calculated skyrmion diameter $d_\textrm{sk}$.}
\begin{tabular}{rrrrrrrrrrr}\hline \hline
\multicolumn{1}{c}{$t_\textrm{Co}$}	&	\multicolumn{1}{c}{$M_\textrm{S}$}		&	\multicolumn{1}{c}{$H_\textrm{K}$}	&	\multicolumn{1}{c}{$K_\textrm{eff}$}	& \multicolumn{1}{c}{$\zeta_0$}	&	\multicolumn{1}{c}{$H_\textrm{c}$} 	&	\multicolumn{1}{c}{$T_\textrm{C}$}	&	\multicolumn{1}{c}{$A$}	&	\multicolumn{1}{c}{$\Delta$}		 &	\multicolumn{1}{c}{$D_\textrm{S}$} & 	\multicolumn{1}{c}{$d_\textrm{sk}$}	\\
\multicolumn{1}{c}{(nm)}	&	\multicolumn{1}{c}{(MA~m${}^{-1}$)}	&	\multicolumn{1}{c}{(MA~m${}^{-1}$)} & \multicolumn{1}{c}{(MJ~m${}^{-3}$)}	&	\multicolumn{1}{c}{(mT)${}^{1/4}$}	&	\multicolumn{1}{c}{(mT)} &	\multicolumn{1}{c}{(K)}	&	\multicolumn{1}{c}{(pJ~m${}^{-1}$)}		&	\multicolumn{1}{c}{(nm)}	&	\multicolumn{1}{c}{(pJ~m${}^{-1}$)}	 &	\multicolumn{1}{c}{(nm)}	\\\hline
0.6	&	1.15	&	0.84	& 	0.60	&	25	&	2.6	& 385 & 1.9 &	1.8	& -0.271	&	44	\\
0.7	&	1.31	&	0.76	&	0.62	& 	31	&	4.1	& 440 & 2.2 &	1.9 & -0.253	&	54	\\
0.8	&	1.37	&	0.70	&	0.60	& 	40	&	6.7	& 479 & 3.0 &	2.2	& -0.169	&	61	\\
0.9	&	1.45	&	0.40	&	0.36	&	78	&	9.4	& 530 & 6.2 &	4.1 & -0.123	&	152	\\
1.0	&	1.48	&	0.27	&	0.25	&	87	&	10.8& -   &11.5 &	6.8	& -0.148	& 	203	\\ \hline\hline
\end{tabular}
\label{Tab:magpar}
\end{table*}

The full approach is validated on a samples series of out-of-plane magnetized //CrOx/Co($t_\textrm{Co}$)/Pt trilayers (Fig.~\ref{Fig:results}a) with varying cobalt layer thicknesses $t_\textrm{Co}$ prepared by sputter deposition. Elementary magnetic characterization is carried out by zero offset anomalous Hall magnetometry\cite{Kosub2015} to extract the anisotropy field $H_\textrm{K}$ from the hysteresis loops (Fig.~\ref{Fig:results}a), supported by SQUID VSM for the determination of $M_\textrm{S}$ and its temperature dependence to extrapolate $T_\textrm{C}$ (Supplementary~\ref{sec:magnetometry}). All samples show perpendicular-to-plane easy axis of magnetization. Towards thinner cobalt layers $M_\textrm{S}$ significantly decreases while the effective anisotropy constant $K_\textrm{eff}$ increases (see  Fig.~\ref{Fig:results}c).\\
Furthermore, we observed the formation of 360~${}^\circ$ DWs or so called \emph{winding pairs}\cite{Hubert1998, Benitez2015} during field driven creep motion as well as a coherent currentinduced DW mediated by spin orbit torques as shown in Fig.~\ref{Fig:Kerr}d and e. From these two qualitative findings the presence of sufficiently strong interface-induced DMI indicative of a homochiral nature of the DWs is concluded.\\
The two parameters $A$ and $D_\textrm{S}$ were determined as described above. As result of the asymmetric domain growth (Fig.~\ref{Fig:Kerr}a,b) we find an almost $t_\textrm{Co}$-invariant DMI strength with a mean value of $D_\textrm{S}=$-0.20$\pm$0.09~pJ/m. This suggests a purely interface induced DMI mechanism for the studied //CrOx/Co($t_\textrm{Co}$)/Pt trilayer system. Since the diameter scaling $\propto (D_\textrm{S}\,t_\textrm{Co})^{-1}$ was not observed (Fig.~\ref{Fig:results}b), the exchange parameter was determined in the framework of magnetic bubble domain theory. As depicted in Fig.~\ref{Fig:results}c increasingly significant deviations as large as an order of magnitude of $A$ from the bulk value were obtained with decreasing $t_\textrm{Co}$. The complete set of all magnetic parameters is summarized in Tab.~\ref{Tab:magpar}.


\section*{Discussion}

The DMI strength is quantified from the asymmetry ratio of DW creep velocities independently from the exchange parameter $A$. The developed low-field analysis of the creep velocities provides a complementary quantification methods to the most commonly employed evaluation of the DMI-induced asymmetric magnon dispersion. 
\\
The DMI strength of $D_\text{S}= - (0.20\pm0.09)$~pJ/m in //CrOx/Co/Pt trilayers is almost independent of the cobalt thickness. This indicates the interface-induced character of the DMI mechanism as previously reported in comparable systems \cite{Belmeguenai2015,Cho2015}. Hence the variation of the cobalt layer thickness is a common approach to tailor the effective DMI constant $D_\text{eff}=D_\text{S}\,t_\text{Co}$. \\
The //Pt(2nm)/Co(1.2nm)/AlOx(2nm) tilayers have large DMI values of $(0.52\pm0.32)$~pJ/m and $(0.69\pm0.22)$~pJ/m  in the as-deposited and annealed state respectively. The opposite sign of the asymmetric DW expansion in //CrOx/Co/Pt and //Pt/Co/AlOx trilayers corresponds to a sign change of the Dzyaloshinskii-vector of the system. In case only the Co/Pt interface contributes to the DMI, the sign change is expected due to the inversion of the stacking order. 

\begin{figure}[!t]
\centering
\includegraphics[width=.4\linewidth]{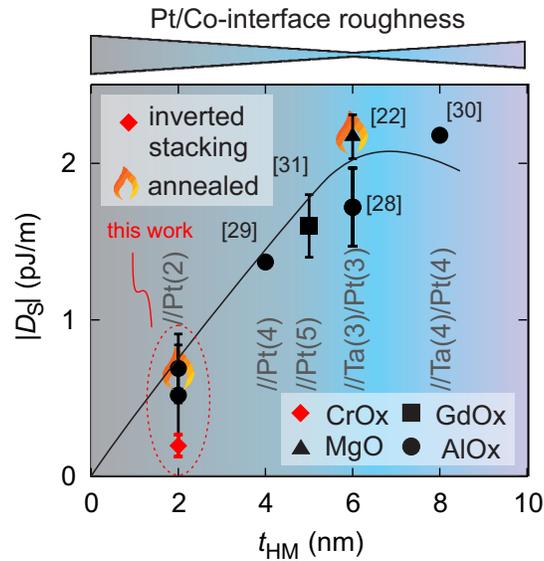}
\caption{Comparison of DMI constants in Pt/Co/metal-oxide trilayers from the literature \cite{Belmeguenai2015, Cho2015, Kim2015a,Vanatka2015, Boulle2016} with the results obtained in //CrOx/Co/Pt trilayers and //Pt/Co/AlOx reference samples. The combined layer thickness $t_\text{HM}$ of Ta-buffer and Pt bottom layer serves as qualitative indicator of the Pt/Co interface roughness \cite{Slepicka2008} reaching a minimum at about 6-8~nm. Plotting $D_\text{S}$ over $t_\text{HM}$ reveals a striking dependence that suggest a predomination of the DMI strength by the Pt/Co interface quality.}
\label{Fig:discussion}
\end{figure}

In following the significant difference of DMI values of the investigated samples to those reported in literature is addressed from a structural point of view. The determined DMI values in the here studied samples are considerably smaller than in other reported //Pt/Co/MO systems \cite{Belmeguenai2015, Cho2015, Kim2015a,Vanatka2015,Di2015a,Boulle2016}.\\
A similar approach to that of Sagasta et al. \cite{Sagasta2016} relating spin-Hall angles to the Pt resistivity to obtain correlations of microstructural properties to the spin-orbit effect strength is applied for the DMI strength.
As indirect indicator of the interface roughness/quality the combined thickness $t_\text{HM}$ of the Ta and Pt in the bottom layer is evaluated (Supplementary~\ref{sec:DMIdisc}). The smoothest interfaces for these systems is expected in a thickness range between 6 and 8~nm. As the behavior in the ultra-thin limit is related to the wetting of metals on the substrate, it is inferred that the heavy metal kind is of subordinate role for the process. As long as the Pt layer is closed (i.e. $t_\text{Pt}>$2~nm) a Ta buffer layer contributes to the smoothening of the layer. Therefore the interface roughness is related to the combined layer thickness of Pt and Ta in this discussion.\\
Fig.~\ref{Fig:discussion} shows that $D_\text{S}$ and $t_\text{HM}$ are also correlated in a striking, almost linear dependence. For this comparison the reference samples with only a 2~nm thick platinum bottom layers were prepared. 
The stack inversion accounts for an additional reduction of the DMI strength, as also found for Pt/Co/MgO trilayers, by a factor of 0.63 \cite{Lee2016} due to a modified growth mode of Co/Pt bilayers on metal oxides, resulting in an even rougher interface. The clear impact of the interface quality on the DMI helps to explain the reduced DMI in the //CrOx/Co/Pt samples as a result of unfavorable growth conditions for Pt on the CrOx/Co buffer layers. The reference samples with Co grown on only 2~nm thick Pt yield are similarly rough but have slightly better interfaces. It is therefore concluded, that the DMI strength is predominated by the Pt/Co interface quality. A cartoon picture of how the roughness may influence the effective DMI constant is presented in (Supplementary \ref{sec:DMIdisc}). According to our comparison of results obtain for DMI constants in stacks with different metal oxides, a DMI contribution from the MO/Co interface is experimentally not evident (Supplementary \ref{sec:DMIdisc}).  \\

The cobalt layer thickness in //CrOx/Co($t_\text{Co}$)/Pt trilayers has a strong impact not only on the saturation magnetization and the effective anisotropy but also on the exchange parameter $A$. After employing the developed approach to determine the DMI constant independently of the exchange parameter, $A$ can be quantified from the size of circular magnetic objects. Judging from the $t_\text{Co}$-scaling of the diameter, the objects are identified as magnetic bubble domains. In the framework of bubble domain theory the exchange parameter can be deduced from the collapse diameter of magnetic bubble domains. For samples satisfying the condition $Q=K_\text{eff}/K_0 > 1.5$ the study of bubble domain stability offers a quasi-static alternative to methods relying on magnetization dynamics \cite{Langer2016, Nembach2015}.\\
As the investigations revealed the exchange parameter in the ultra thin Co layer can be diminished up to an order of magnitude compared to the cobalt bulk exchange value $A^b=$16~pJ/m. This trend is supported by the approximative estimation of $A$ from the extrapolated Curie temperature. A similar reduction of $A$ was reported in a comparable system \cite{Nembach2015}. The reduction is attributed to a micro-structural origin, e.g. an increasing number of grain boundaries with decreasing cobalt layer thickness in the ultrafinely polycrystalline samples.\\
With the significantly diminished $A$ the expected domain wall width $\Delta$ as well as the exchange length $l_\text{ex}$ can be smaller than 2~nm responsible for the strong susceptibility of DWs to small pinning sites as e.g. grain boundaries. 

With the determined DMI strength all samples safely remain in the limit of a collinear ferromagnetic ground-state. Due to the large anisotropy the effective DMI constants are at most 33~\% of the theoretical limit or the critical DMI constant $D_\text{c} = 4\,\sqrt{A\,K_\text{eff}}/\pi$ \cite{Bogdanov1994}, for the formation of modulated spiral and skyrmion phases.\\
The DW energy is reduced by 1-6~\% in the CrOx/Co/Pt samples due to the DMI. However, it is sufficient to induce modifications to the DW providing homochiral and Néel-type properties. Consequently the DWs obtain homochiral and Néel-type properties that are responsible for the formation of winding pairs and coherent current induced DW motion. It is therefore concluded that the magnetic bubble domains are confined by a homochiral DW. \\
With the complete set of micro-magnetic parameters the skyrmion diameter can be calculated. For the studied sample series the expected skyrmion diameter $d_\text{Sy}$ lies in a range between 40 and 200~nm. An upper limit for the $d_\text{Sy}$ is given by considering the bulk exchange value in the micromagnetic simulations. \\


\section*{Methods}

\indent \textbf{Sample preparation.} A sample series of perpendicularly magnetized //CrOx(5~nm)/Co($t_\textrm{Co}$)/Pt(2~nm) was magnetron-sputtered at room temperature on thermally oxidized silicon substrates with cobalt layers of varying thickness $t_\textrm{Co}$ and an argon pressure of 10${}^{-3}$~mbar. The Pt top layer thickness of only 2~nm is sufficient to serve as a oxidation barrier providing large optical transmittance and low magnetoresistive shunting. The thin amorphous CoOx buffer layer is nonmagnetic (Supplementary). Optical lithography and subsequent Ar-Ion-Beam etching were used for patterning.\\

\textbf{DMI quantification.} For the quantification of the DMI constant we used wide field Kerr microscopy with maximum fields up to $H_\textrm{x}=$250~mT and $H_\textrm{x}=$20~mT. The latter is provided by a small air-core coil, which is driven by a Keysight Sourcemeter providing field pulses of defined width down to 2~ms. The DW velocities are determined from the traveled distance $s$ measured from differential Kerr-images taken from the initial and final states after application of an $H_\textrm{z}$ current pulse with the width of $\Delta t$. Great caution was given to avoid a misalignment of the sample with respect to the hard axis field, that causes an additional $H_\textrm{z}$ component. To cancel any remaining effect of a possible misalignment the respective velocities for $\uparrow\downarrow$ and $\downarrow\uparrow$ DW from both up and down domains were averaged (Supplementary~\ref{sec:creep}).\\
The quasi-static modification of the DW energy by the $H_\text{DM}$ and $H_\text{x}$ can be accounted for by the introduction of $\zeta=\zeta_0\,\left[\sigma^\pm(H_\text{x},\,H_\text{DM})/\sigma_0\right]^{\mu}$ in the creep law\cite{Je2013,Hrabec2014}, with $\zeta_0 = \zeta(H_\text{x})$. The DW energy $\sigma(H_\text{x},\,H_\text{DM})$ in the presence of both fields is reduced from the Bloch wall energy $\sigma_0=4\sqrt{A\,K_0}$ to the following ratio
\begin{equation}
\frac{\sigma^\pm(H_\text{x},H_\text{z})}{\sigma_0}=1-\frac{\pi^2\,\Delta\,M_\text{S}^2}{8\,N_\text{x}\,K_\text{D}\,\sigma_0}\,\mu_0^2 (\pm H_\text{x}+H_\text{DM})^2
\end{equation}
with the shape anisotropy energy scaled by the demagnetization factor $N_\text{x}=4\,\ln (2)\,t_\text{Co}/\Delta$ of the DW. This expression is valid as long as the field combination of $|H_\text{x}+H_\text{DM}|$ is smaller than the transition field from Bloch to N\'eel-Wall $H_{B\rightarrow N}= \frac{2}{\pi}\,N_\text{x}\,M_\text{S}$. 
It is further assumed, that the pinning parameters $\xi$ and $\delta$ entering the quantity $\zeta_0$ are $H_\text{x}$-invariant\cite{Je2013}.\\
For DWs segments situated in an (anti-)parallel ($H_\text{x},H_\text{DM}$)-field configuration one can write the following system of equations 
\begin{equation}
\begin{cases}
\ln v^{+} = \ln v_0 - \zeta_0\,\frac{\sigma^{+}(H_\text{x},H_\text{DM})}{\sigma_0}\\
\ln v^{-} = \ln v_0 - \zeta_0\,\frac{\sigma^{-}(H_\text{x},H_\text{DM})}{\sigma_0}.
\end{cases}
\label{eq:systeq}
\end{equation}
In Eq.~(\ref{eq:systeq}) the creep parameter $\ln v_0$, that can be determined less accurately due to extrapolation (Supplementary~\ref{sec:creep}), is eliminated, which leads to the following expression
\begin{equation}
\ln\left(\frac{v^+}{v^-}\right)\,\frac{(\,\mu_0\,H_\text{z})^{1/4}}{\zeta_0}=\left(\frac{\sigma^+(H_\text{x},H_\text{DM})}{\sigma_0}\right)^{1/4}-\left(\frac{\sigma^-(H_\text{x},H_\text{DM})}{\sigma_0}\right)^{1/4}
\label{eq:num}
\end{equation}
Eq.~(\ref{eq:num}) can be solved numerically for $H_\text{DM}$ (Supplementary~\ref{sec:creep}). 
With the following expansion that is valid when DW energies are only slightly changed by total effective field
\begin{equation}
\left(\frac{\sigma^\pm(H_\text{x},H_\text{z})}{\sigma_0}\right)^{1/4} = 1- \frac{1}{4}\,\frac{\pi^2\,\Delta\,M_\text{S}^2}{8\,N_\text{x}\,K_D\,\sigma_0}\,\mu_0^2 (H_x^2+H_{DM}^2\pm 2\,H_\text{x}\,H_\text{DM}),
\end{equation}
an analytical relation between $H_\text{DM}$ and the velocity ratio $(v^+/v^-)(H_\text{x},H_\text{z})$ in applied fields can be derived from Eq.~(\ref{eq:num}), that reads
\begin{equation}
\mu_0H_\text{DM} = \frac{8\,N_\text{x}\,K_\text{D}\,\sigma_0}{\mu_0\,H_\text{x}\,\pi^2\,\Delta\,M_\text{S}^2}\,\ln\left(\frac{v^+}{v^-}\right)\,\frac{(\mu_0\,H_\text{z})^{1/4}}{\zeta_0}.
\label{eq:HDM-ADG}
\end{equation}
Note, that the creep parameter $\zeta_0$ depends on the exchange constant $A$ (and all the  other micromagnetic and microstructural parameters).  Here, $\zeta_0$ is directly and independently  determined from the experiments in the perpendicular field  $H_\text{x}$ ($H_\text{x} = 0$). Inserting  this parameter now in expressions (\ref{eq:HDM-ADG}) and (\ref{eq:DS-ADG}) yields the DMI strength from the second set of experiments with  $H_\text{x} \neq 0$, but small. 
Eq.~(\ref{eq:HDM-ADG}) can then be rescaled to the DMI strength according to  $H_\textrm{DM}=D_\textrm{S}/(\mu_0\,M_\textrm{S}\,\sqrt{A/K_\textrm{eff}}\,t_\textrm{Co})$
\begin{equation}
D_\text{S}  = \frac{64\,\ln(2)}{\pi^2}\,\ln\left(\frac{v^+(H_\text{x},H_\text{z})}{v^-(H_\text{x},H_\text{z})}\right)\,\frac{(\mu_0\,H_\text{z})^{1/4}}{\zeta_0\,H_\text{x}} \,M_\text{S}\,K_\text{eff}.
\label{eq:DS-ADG}
\end{equation}
With this relation the interface-induced DMI constant can be directly calculated from the logarithmic asymmetry ratio of the DW-velocities $\ln(v^{+}/v^{-})$ obtained at one field combination of $H_\text{x}$ and $H_\text{z}$, the creep parameter $\zeta_0$ and the two elementary magnetic parameters $M_\text{S}$ and $K_\text{eff}$. It is emphasized that this expression does not explicitly depend on the exchange parameter  $A$ and is applicable only for small $H_\text{x}$.\\

\textbf{Determination of the exchange parameter.} Patterning the samples into stripes of 60~$\mu$m width with a constriction to 3~$\mu$m in the scheme of Ref.~\cite{Jiang2015} allows for the creation of circular magnetic objects. By simultaneous application of current pulses and small easy axis field ($H_\textrm{z}$), homochiral DWs are pushed along the stripe by spin orbit torques.\cite{Emori2013,Ryu2014} A homochiral circular object is formed due to the contraction of a domain channel in divergent currents and the expulsion from the constriction area.\\
In the studied system a scaling of the diameter of the generated objects $\propto (D_\textrm{S}\,t_\textrm{Co})^{-1}$ as expected for skyrmions is not observed. Hence, an analysis in the framework of isolated magnetic bubble domains rather than skyrmion theory is required. The typical diameter of these bubble domains is of the order the collapse diameter $D_\textrm{bc}$ of bubble domains (Supplementary~\ref{sec:scaling}). 
According to bubble theory,\cite{Hubert1998} valid for $Q = K_\textrm{I}/K_\textrm{D} > 1.5$ with $K_\textrm{I}= K_\textrm{eff} + K_\textrm{D}$, the reduced collapse diameter $d_\textrm{bc}=D_\textrm{bc}/t_\textrm{Co}$  can be derived from the ratio of DW energy $\sigma_{DW}$ to shape anisotropy $K_\textrm{D}= \mu_0\,M_\textrm{S}/2 $ satisfying the condition: 
\begin{equation}
\lambda_c= \sigma_{DW}/(2\,K_\textrm{D}\,t_{Co}) = S_\textrm{bc}(d_\textrm{bc}) = \frac{2}{\pi}\left[ d_\textrm{bc}^2\,(1-\mathrm{E}(u^2)/u)+u\,\mathrm{K}(u^2)  \right],
\label{eq:AQ}
\end{equation}
with the complete elliptic integrals $ \mathrm{E}(u)$ and $\mathrm{K}(u)$ as function of $u = \sqrt{d^2/(1+d^2)}$
\begin{equation}
\mathrm{E}(u) = \int_0^{\pi/2}\sqrt{1-u\sin^2\alpha} \mathrm{d}\alpha, \quad  \mathrm{K}(u) = \int_0^{\pi/2}\mathrm{d}\alpha/\sqrt{1-u\sin^2\alpha} .
\end{equation} 
We point out that in the presence of DMI the energy of DWs with the favored chirality reads $\sigma_{DW} = 4\sqrt{A\,K_\textrm{eff}}- \pi\,D_\textrm{S}/t_\textrm{Co}$. 
Solving equation~\ref{eq:AQ} for the exchange parameter $A$, as the only unknown material parameter remaining, yields expression~(\ref{eq:Aquant}
Furthermore, the correct experimental conditions can be verified by calculating the expected collapse field $H_\textrm{bc}=M_\textrm{S}\,t_\textrm{Co}\,[F(d_\textrm{bc})-S_\textrm{bc}(d_\textrm{bc})]/d_\textrm{bc}$ from the condition of the force function 
\begin{equation}
F(d) = -\frac{2}{\pi}\,d^2 \left[1-\mathrm{E}(u^2)/u\right] = \lambda_c+\frac{H_\textrm{bc}}{M_\textrm{S}}\,d
\end{equation}
proving that the experimental fields were just slightly below the theoretical collapse field (Supplementary~\ref{sec:scaling}).
Note that in case of the largest cobalt thickness deviations might occur due to the violation of the condition $Q = K_\textrm{I}/K_\textrm{D} > 1.5$.\\

\textbf{Calculation of skyrmion diameters.} For the analysis of skyrmion diameters, we use the micromagnetic framework with the continuous description of Dzyaloshinskii-Moriya interaction~\cite{Rohart2013,Fert2013}. As a model, we consider a thin ferromagnetic infinite plate of thickness $h$ along the $\vec{\hat{z}}$-axis. Taking into account that thickness of ferromagnetic layer is smaller than characteristic magnetic length, we consider a uniform average values of magnetic parameters along the $\vec{\hat{z}}$-axis. Also, we assume that the magnetostatic interaction, which is always present in the system, can be reduced to the easy-surface anisotropy, which results to the appearance of effective anisotropy coefficient $K_{\textrm{eff}}$. Due to this, the total micromagnetic energy density of our system will have a following form
	 \begin{equation} \label{eq:Total_energy_density_cartesian}
    \mathcal{E}= A \left[ (\nabla m_{x})^2 + (\nabla m_{y})^2 + (\nabla m_{z})^2 \right] - D_{\textrm{S}} \left[ m_{n} \nabla \cdot \vec{m} - (\vec{m} \cdot \nabla) m_{n} \right] - K_{\textrm{eff}} \, (\vec{m} \cdot \vec{n})^2 - \vec{M} \cdot \vec{B},
	 \end{equation}
	where we take into account the exact form of interface-induced DMI term, $\vec{n}$ is the unity vector directed perpendicular to the interface surface along the $\vec{\hat{z}}$-axis. For the case of simplicity, we introduce cylindrical coordinate system with $\vec{r} = (r \cos \chi, r \sin \chi, z)$ and use angular parametrization for the reduced magnetization vector $\vec{m} = \vec{M}/M_{S}=(\cos \phi \sin \theta, \sin \phi \sin \theta, \cos \theta)$. In this case, the total energy density \eqref{eq:Total_energy_density_cartesian} will have the following form: 
	\begin{equation} \label{eq:Total_energy_density_cylindrical}
    \begin{split}
  		\mathcal{E} = A \left\lbrace \left( \dfrac{\partial \theta}{\partial r} \right)^2 + \dfrac{1}{r^2} \left( \dfrac{\partial \theta}{\partial \chi} \right)^2 + \sin^2 \theta \left[ \left( \dfrac{\partial \phi}{\partial r} \right)^2 + \dfrac{1}{r^2} \left( \dfrac{\partial \phi}{\partial \chi} \right)^2 \right] \right \rbrace - K_{\textrm{eff}} \, \cos^2 \theta - M_{\textrm{S}} B \, \cos \theta - \\
		- D_{\textrm{S}}  \left\lbrace  \sin (\phi - \chi) \left[ \dfrac{1}{r} \dfrac{\partial \theta}{\partial \chi} - \sin \theta \cos \theta \dfrac{\partial \phi}{\partial r} \right] + \cos (\phi - \chi) \left[ \dfrac{\partial \theta}{\partial r} + \dfrac{\sin \theta \cos \theta}{r} \dfrac{\partial \phi}{\partial \chi} \right]  \right \rbrace. 
            \end{split}
	\end{equation}
	We are looking for azimuthally symmetric solutions, which represent either skyrmion or skyrmion-bubble textures~\cite{Butenko2010,Leonov2016}:
	\begin{equation} \label{eq:Simple_solutions}
	\theta = \theta(r), \qquad \phi = \phi(\chi),
	\end{equation}
	where $\phi(\chi)$ is a linear function with respect to $\chi$. This allows us to simplify the expression \eqref{eq:Total_energy_density_cylindrical} and derive the total energy of a chiral axisymmetric magnetic texture:
	\begin{equation} \label{eq:Total_energy_cylindrical_simplified}
		E = 2 \pi \int_0^{\infty} \mathrm{d} r \left \lbrace A \left(\theta_r^2 + \dfrac{1}{r^2} \sin^2 \theta \right)-D_{\textrm{S}} \left(\theta_r+\dfrac{1}{2 r} \sin (2 \theta)\right)-K_{\textrm{eff}} \cos^2 \theta - B \, M_\textrm{S} \cos \theta \right \rbrace.
	\end{equation}
	
	The Euler-Lagrange equation for the total energy functional~\eqref{eq:Total_energy_cylindrical_simplified} has the following form,
	\begin{equation} \label{eq:theta_equation}
		A \left[ \theta_{rr} + \dfrac{\theta_r}{r} - \dfrac{\sin (2\theta)}{2 r^2} \right] - D_{\textrm{S}} \dfrac{\sin^2 \theta}{r} - \dfrac{1}{2} K \sin(2 \theta) - \dfrac{1}{2} B \, M_\textrm{S} \sin \theta =0,
	\end{equation}
	with boundary conditions for isolated axisymmetric skyrmion:
	\begin{equation} \label{eq:boundary_conditions}
		\theta(0) = 0, \qquad \theta(\pi) = \pi.
	\end{equation}
	
	The boundary value problem \eqref{eq:theta_equation} and \eqref{eq:boundary_conditions} can be solved numerically by using finite-difference method. Usually the function $\theta(r)$ has spike-like shape in the vicinity of point $r=0$ and decays exponentially at hight distances from the skyrmion center. The characteristic size of a localized magnetization profile $\theta(r)$ is usually defined as~\cite{Butenko2010,Roessler2006,Leonov2016} 
	\begin{equation}
		r_{\textrm{sk}} = r_0 - \theta_0 \left( \dfrac{\mathrm{d} \theta}{\mathrm{d} r}\right)^{-1}_{r=r_0},
	\end{equation}
	where $(r_0, \theta_0)$ is the inflection point of the profile $\theta(r)$.

\bibliography{references}

\section*{Acknowledgements}

We thank C. Krien (IFW Dresden) for sputter deposition of the metal multilayers, D. Stein (IFW Dresden) for Kerr microscopy measurements, G. Rane (IFW Dresden), A. Scholz and J. Grenzer (HZDR) for XRR measurements, C. Xu (HZDR) for SQUID-VSM measurements,  S. Facsko for performing and analyzing the AES measurements as well as T. Schneider (HZDR) for the help with MuMax simulation. Support by the Structural Characterization Facilities at IFW Dresden and IBC of the HZDR is gratefully acknowledged. This work was funded in part by the
European Research Council under the European Union's Seventh Framework Programme
(FP7/2007-2013)/ERC grant agreement no. 306277 and the European Union Future and
Emerging Technologies Programme (FET-Open Grant No. 618083).

\section*{Author contributions statement}

M.K. and T.K. set up the magneto-transport measurements. A.K. set up the environment and code for the micromagnetic calculations.
M.K. carried out the elementary magnetometry and all Kerr microscopy experiments as well as the corresponding data analysis. H.F. wrote an analysis script for the evaluation of domain wall velocities. U.K.R. gave supporting theoretical background. R.S. provided support for Kerr microscopy measurements. O.V. calculated the skyrmion diameters. M.K., T.K. and D.M. created the graphics and M.K., U.K.R, T.K., A.K.and D.M. wrote the manuscript with comments from all authors. D.M., O.G.S., J.L. and J.F. supervised the project.

\section*{Additional information}

To include, in this order: \textbf{Accession codes} (where applicable); \textbf{Competing financial interests} (mandatory statement). 

The corresponding author is responsible for submitting a \href{http://www.nature.com/srep/policies/index.html#competing}{competing financial interests statement} on behalf of all authors of the paper. This statement must be included in the submitted article file.

\clearpage
\onecolumn
\normalsize
\section*{Supplementary Information}

\setcounter{figure}{0}
\renewcommand{\thefigure}{S\arabic{figure}} 

\setcounter{table}{0}
\renewcommand{\thetable}{S\arabic{table}} 

\renewcommand{\thesection}{S\arabic{section}} 

\setcounter{equation}{0}
\renewcommand{\theequation}{S\arabic{equation}} 

\section{Magnetometry}
\label{sec:magnetometry}

\begin{figure}[!b]
\centering
\includegraphics[width=1\linewidth]{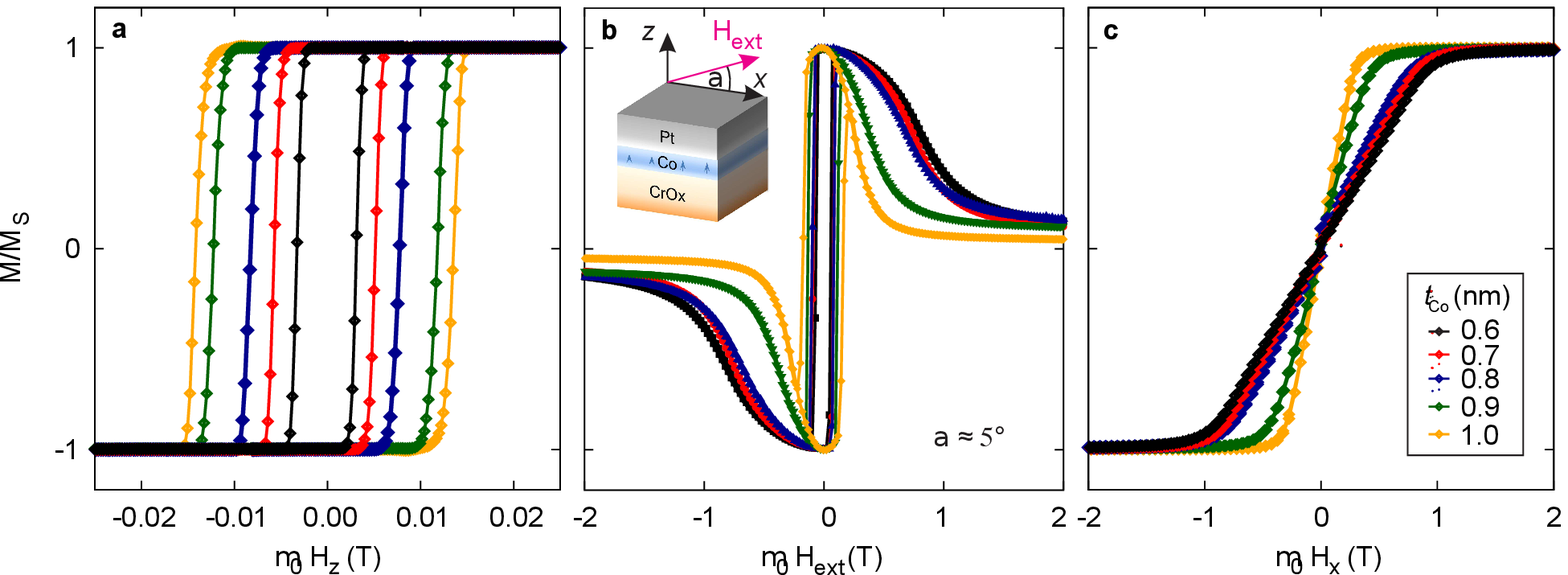}
\caption[Magnetic field-dependent properties of //CrOx/Co/Pt trilayers]{Normalized AHE measurements with the field applied \textbf{(a)} parallel to the surface normal ($\mathbf{\hat{z}}$, $\alpha =90^\circ$) and  \textbf{(b)} under an angle $\alpha$ of about  $5^\circ$  with respect to the $\mathbf{\hat{x}}$ axis (see sketch). The hysteresis loops of the //CrOx/Co($t_\text{Co}$)/Pt trilayers in \textbf{(a)} easy  and  \textbf{(c)} hard axis direction of the magnetization are obtained, after a rescaling of the curves in panel \textbf{(b)} according to Eq.~(\ref{eq:harescale}) }
\label{Fig:magnetometry}
\end{figure}

The full hysteresis loops (Fig.~\ref{Fig:magnetometry}a,c) are obtained from anomalous Hall effect (AHE) measurements. As the hysteresis loops measured by vibrating sample magnetometry (VSM) are susceptible to magnetic contaminations adding an unknown paramagnetic offset to the hysteresis loops, only the remanent magnetization values are taken into account.  
Apart from a small linear contribution of 10~m$\Omega$/T attributed to the ordinary Hall effect of the platinum layer, all studied samples show full remanence and the determined magnetization value is a direct measure of the saturation magnetization $M_\text{S}$ for these samples under this condition. Therefore it is justified to use the remanent magnetization in easy axis direction as the saturation magnetization value of the samples.\\
To calculate the hard axis loops from Hall-loops measured in tilted geometry (see Fig.~\ref{Fig:magnetometry}b), the following geometrical relations are employed. In the tilted geometry the field component, that is effectively applied in the direction of the hard axis direction of magnetization, is equal to the projection of the external field vector onto the film plane, $H_\text{x}=H_\text{ext}\cdot \cos\alpha$. The angle $\alpha=\arccos(H_\text{c}/H_\text{c}^\text{t})$ is calculated from the ratio of the respective switching fields of easy axis and tilted hysteresis loop. Given that the samples exhibit full remanence the inplane magnetization component at any field point can be deduced from the following geometrical relation \\
\begin{equation}
 \quad M_\text{x}(H_\text{x})=\sin\left[\arccos\left(\frac{R^\text{H}(H_\text{x})}{R^\text{H}(H_\text{x}=0)}\right)\right], 
\label{eq:harescale}
\end{equation} 
with the ratio of the field-dependent Hall resistance $R^\text{H}(H_\text{x})$ to its value obtained at zero field in the fully magnetized remanent state.\\
The anisotropy field $\mu_0H_\text{K}$ is deduced from the hard axis loop since it is equal to the integral of $1-|M(\mu_0H_\text{ext})/M_\text{S}|$ according to the definition of the effective anisotropy energy constant $K_\text{eff}=\mu_0\,M_\text{S}\,H_\text{K}/2$.\\

\begin{figure}[!t]
\includegraphics[width=1\linewidth]{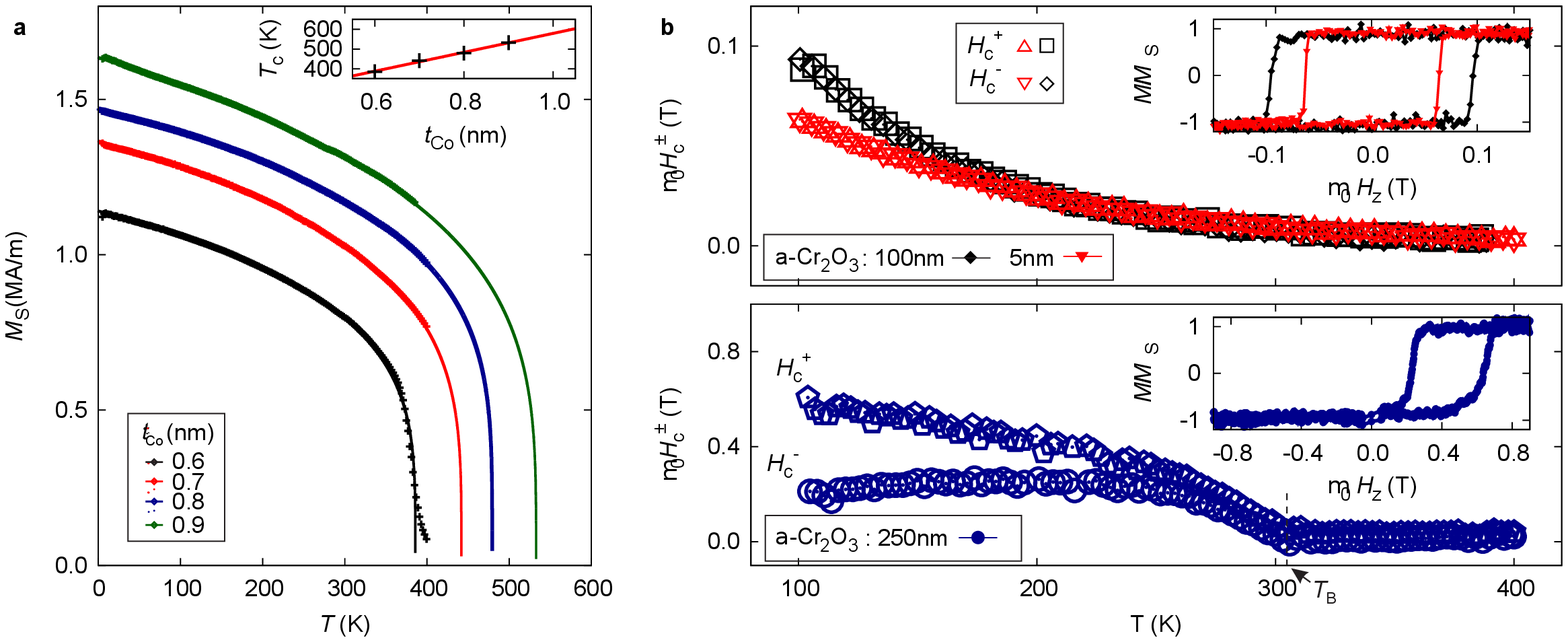}
\caption[Temperature-dependent magnetic properties of //CrOx/Co/Pt trilayers]{\textbf{(a)} Temperature dependence of the magnetization $M_\text{S}$ allows for the extrapolation of the Curie temperatures. \textbf{(b)} The study of the temperature depend coercive fields  $H^\pm_\text{c}$ of different Cr${}_2$O${}_3$ seed layers reveals possible antiferromagnetic properties. Amorphous a-Cr${}_2$O${}_3$ seed layer with thicknesses of 5~nm and 100~nm show symmetric switching behavior, i.e. the left $H_\text{c}^-$ and right $H_\text{c}^+$ coercive fields are equal. For crystalline $\alpha$-Cr${}_2$O${}_3$ seed layers sizable loop shifts of the easy axis hysteresis loop by an exchange bias field $H_\text{EB}$ up to the blocking temperature $T_\text{B}$ of 305~K clearly indicated anti-ferromagnetic properties.}
\label{Fig:Hc(T)}
\end{figure}

The following temperature dependent properties were obtained by vibrating sample magnetometry. By fitting the temperature dependent magnetization curves (see Fig.~\ref{Fig:Hc(T)}a) to the following expression
\begin{equation}
M(T)= M_\text{S}\,[(T_\text{C}-T)/T_\text{C}]^\beta,
\label{eq:M(T)}
\end{equation}
the Curie temperatures $T_\text{C}$ are extrapolated. A value of 0.75 for the critical exponent $\beta$ is determined from the fit for all samples. The analysis suggests a strong reduction of $T_\text{C}$ from the bulk value of $T_\text{C}^b$=1394~K in the thin films.\\

During cool-down from 400~K to 100~K the coercive fields $H_\text{c}$ of three samples with different Cr${}_2$O${}_3$ seed layers were tracked in VSM by constantly sweeping the field. Fig.~\ref{Fig:Hc(T)}b shows that left and right switching field branches are symmetric in the whole temperature range not only for the sample with 5~nm thick amorphous a-Cr${}_2$O${}_3$ but also in the case of the much larger thickness of 100~nm. This clearly proves the absence of anti-ferromagnetic properties in the a-Cr${}_2$O${}_3$ layers that otherwise would induce an exchange bias (EB) effect.\\
In contrast, the EB effect is very obvious in the reference sample with crystalline $\alpha$-Cr${}_2$O${}_3$. The hysteresis loop is shifted up to 400~mT below the blocking temperature of about 305~K.\\
The absence of magnetoelectric properties in a-Cr${}_2$O${}_3$ layers down to temperatures of 100~K confirms the former finding that a strong perpendicular magnetic anisotropy (PMA) induced by Cr${}_2$O${}_3$  does not originate from the perpendicular exchange coupling effect at crystalline $\alpha$-Cr${}_2$O${}_3$/FM interface \cite{Nozaki2013}.\\
The usage of amorphous chromium oxide (a-Cr${}_2$O${}_3$) for spin-orbit effect studies has hitherto been neglected, due to its possible magneto-electric properties and related effects. The strong PMA induced by Cr${}_2$O${}_3$ allows the variation of the FM thickness in a reasonably large range. The aim is to preserve perpendicularly magnetized films from the same sandwiched system over a significant large range. Thus, the possibility to adjust the magnetic properties is provided and the influence of the interfaces can be studied.

\newpage

\section{Field-driven DW creep motion}
\label{sec:creep}

The nucleation field distribution is very sharp at small $t_\text{Co}$ and broadens for larger thicknesses accompanied by an increase of nucleation sites. For similar Pt/Co/Pt systems an asymmetric nucleation for opposite initial magnetization saturation states at defects has been reported \cite{Iunin2007}, which might stem from the recently discussed step-edge-induced DMI \cite{Pizzini2014}. \\
In a certain field range, that is different for each sample, DW motion is observed such that it does not develop overhanging wall sections. For larger $t_\text{Co}$ this field range starts to overlap with the nucleation field distribution, so that in at higher fields it can not be discerned, whether a sample area remagnetized due to nucleation or wall propagation.\\
The field-driven DW propagation is classified into different propagation regimes \cite{Metaxas2007,Gorchon2014}. With successively increasing external magnetic field the creep, thermally assisted flux flow and depinning regimes are observed until the flow motion regime is asymptotically reached. While the highly non-linear response at low fields is well understood in terms of the creep law, the regimes close to the depinning are both experimentally and theoretically less clear according to  Ref.~\cite{Gorchon2014}. 
Also the typical field ranges of the flow regime are hard to realize in experiment as they have to be applied as well-defined short-time large-amplitude field pulses. For these two reasons the following investigations are performed in the creep regime. 

\begin{figure}[!b]
\centering
\includegraphics[width=.8\linewidth]{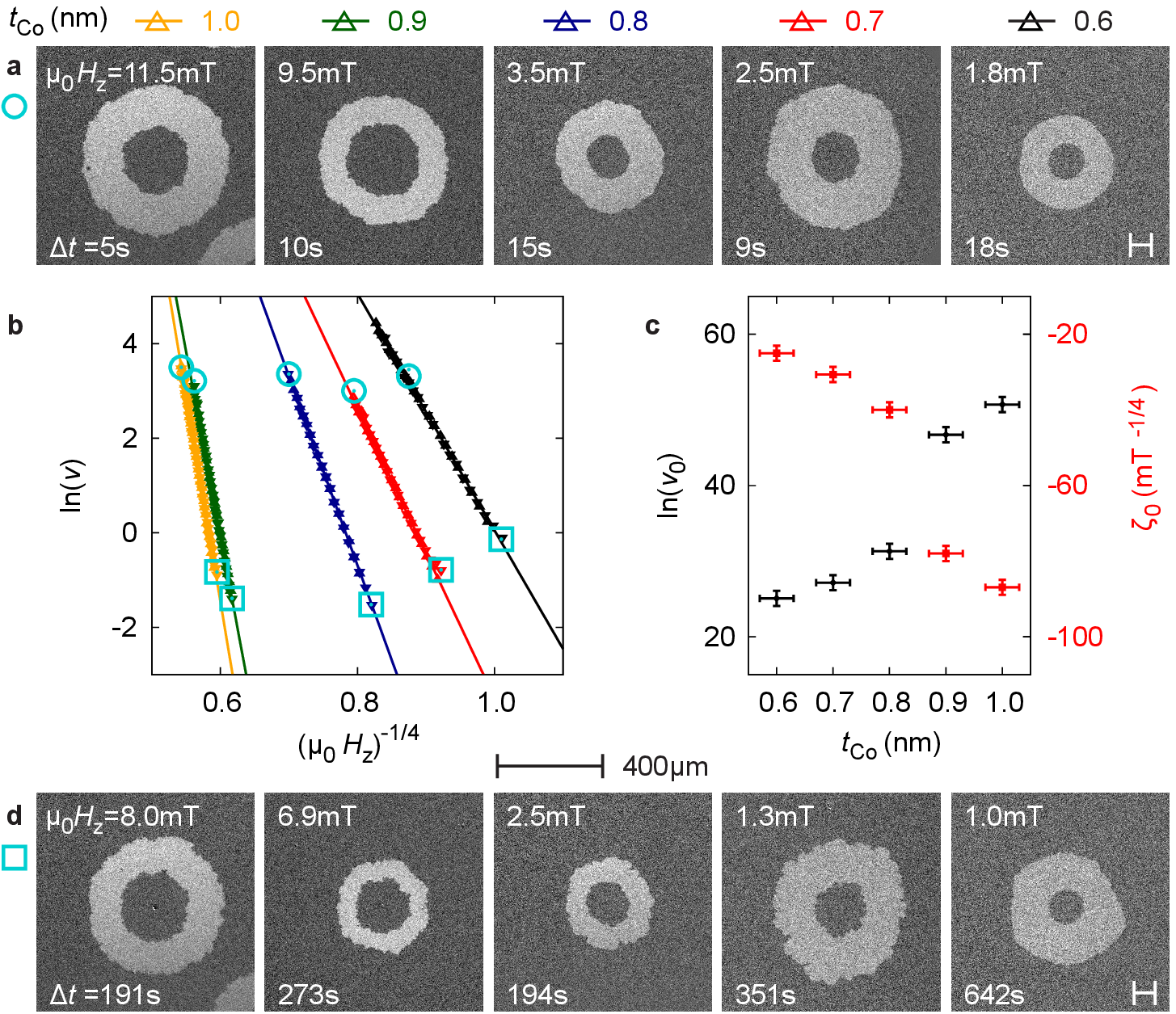}
\caption[Domain wall creep motion in //CrOx/Co/Pt trilayers]{Panel \textbf{(a)} and  \textbf{(d)} show differential Kerr images of the circular field-driven domain expansion in //CrOx(5)/Co($t_\text{Co}$)/Pt(2) trilayers at high and low external magnetic fields respectively for different cobalt thicknesses $t_\text{Co}$. \textbf{(b)} The DW velocities in the applied field regime obey the creep law. The blue square or circle allocates the respective Kerr images to the point in the graph. \textbf{(c)} the parameters $\ln v_0$ and $\zeta_0$ obtained from a linear fit.}
\label{Fig:creep}
\end{figure}

Under small magnetic fields $H_\text{z}$ applied along the easy axis of magnetization, the DW velocities obey the creep law \cite{Lemerle1998,Metaxas2007}
\begin{equation}
v(H_\text{z})=v_0\, \exp \left[-\zeta \,(\mu_0 H_\text{z})^{-\mu}\right)], \quad \mu = 1/4
\label{eq:creep}
\end{equation}
with the value of the critical exponent $\mu$ exactly $1/4$. The two creep law parameters  $\ln v_0$ and $\zeta$ can be deduced from the linear fit in the $\ln(v)\left((\mu_0\,H_\text{z})^{-1/4}\right)$ representation of the data (Fig.~\ref{Fig:creep}b).\\
The parameters $v_0$ and $\zeta$ are functions of the pinning properties of the sample \cite{Lemerle1998} determined by the correlation length $\xi$ and the pinning strength $\delta$ of the disorder 
\begin{equation}
v_0 = \xi\,f_0, \quad \zeta =\frac{U_\text{c}}{k_\text{B}\,T} \,(\mu_0\,H_\text{crit})^\mu
\label{eq:zeta0}
\end{equation} 
with the attempt frequency $f_0$, pinning barrier $U_\text{c}(\xi,\delta)$ and critical field $H_\text{crit}(\xi,\delta)$.\\
A systematic behavior of the scaling parameters as function of the cobalt thickness $t_\text{Co}$ is obtained that is similar to other intrinsic magnetic parameters. The thicker $t_\text{Co}$ and the lower the creep velocity, the more disrupted is the circular shape of the domains. The origin of this DW morphology differences lies in the increasing amount of pinning sites \cite{Metaxas2007}.\\
Both the nucleation field of the investigated pinning site and the creep motion behavior are symmetric under magnetization reversal. The observation of macroscopic DW creep inherently measures an average DW dynamics that neglects local lateral contributions due to film roughness or the granular structure.  This average effectively describes  an asymmetric film, where only the polar $\mathbf{\hat{z}}$-axis is distinguished with an effective $C_{\text{v}\infty}$-symmetry. Locally the defects naturally break this polar simple structure and contribute to the transient pinning of the wall.  Since the domain shape remains circular, there is no evidence for an effective lower symmetry and additional DMI contributions that break the $C_{\text{v}\infty}$-symmetry, which allows the description in terms of the theoretical assumption of the symmetry breaking from the $\mathbf{\hat{z}}$-interface.

\begin{figure}[!p]
\centering
\includegraphics[width=.8\linewidth]{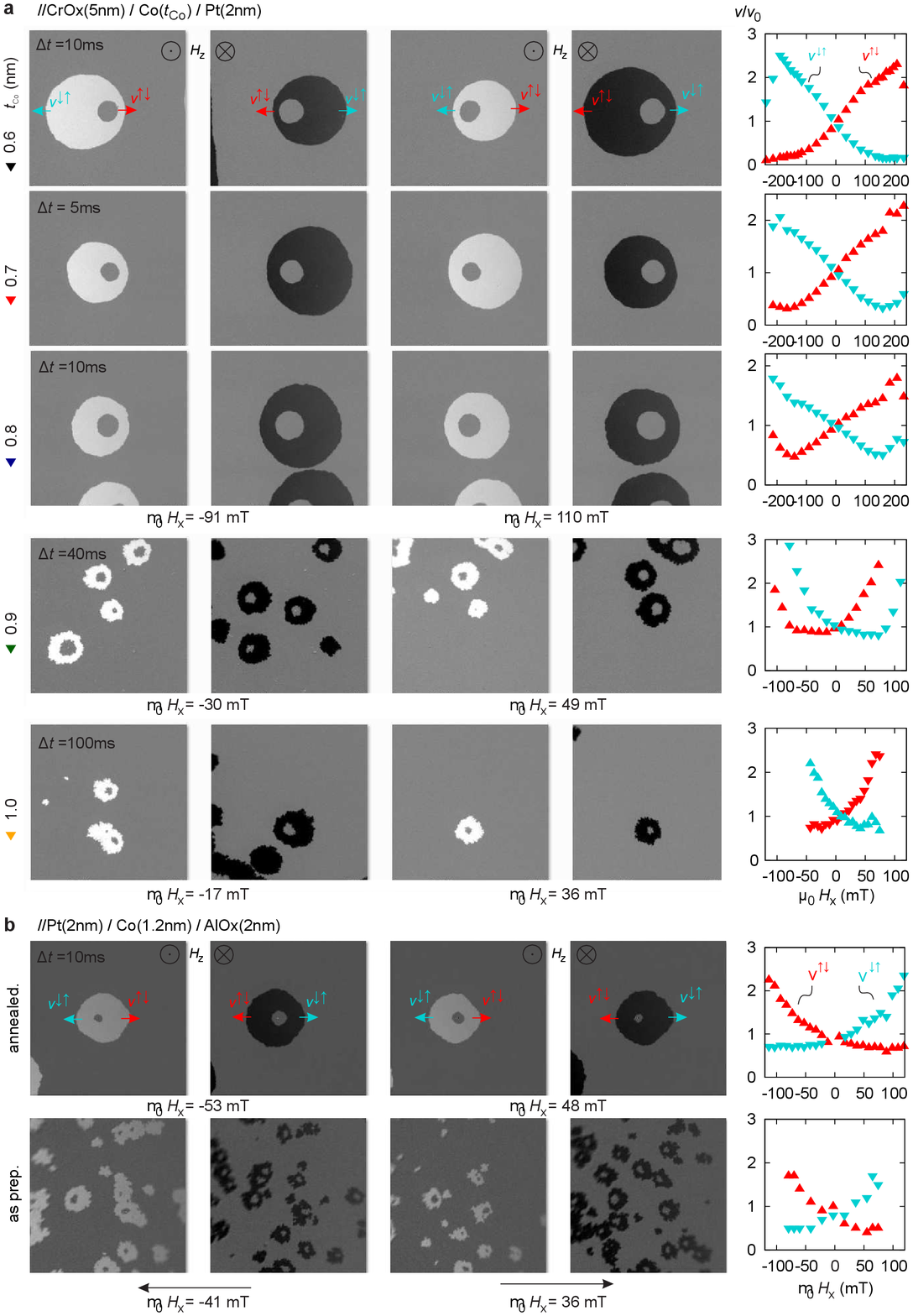}
\caption[]{Caption next page.}
\label{Fig:hrabec}
\addtocounter{figure}{-1}
\end{figure}

\begin{figure}[!p]
\centering
\caption[Asymmetric domain growth in //CrOx/Co/Pt trilayers]{Field-driven asymmetric domain expansion in //CrOx(5)/Co($t_\text{Co}$)/Pt(2) and //Pt(2)/Co(1.2)/AlOx(2) trilayers. The differential Kerr microscopy images taken from the initial and final state before and after a $H_\text{z}$-pulse in combination with an applied $H_\text{x}$-field show an asymmetric distortion. The DW creep velocity around the circumference of the domain is altered according to the the relative orientation of the external field $H_\text{x}$ and the DMI-field $H_\text{DM}$ in the DW. The DW velocities of $\uparrow\downarrow$ and $\downarrow\uparrow$ DWs moving in the anti-/parallel alignment of $H_\text{DM}$ and $H_\text{x}$ are plotted as function of $H_\text{x}$.}
\vspace{10pt}
\includegraphics[width=.4\linewidth]{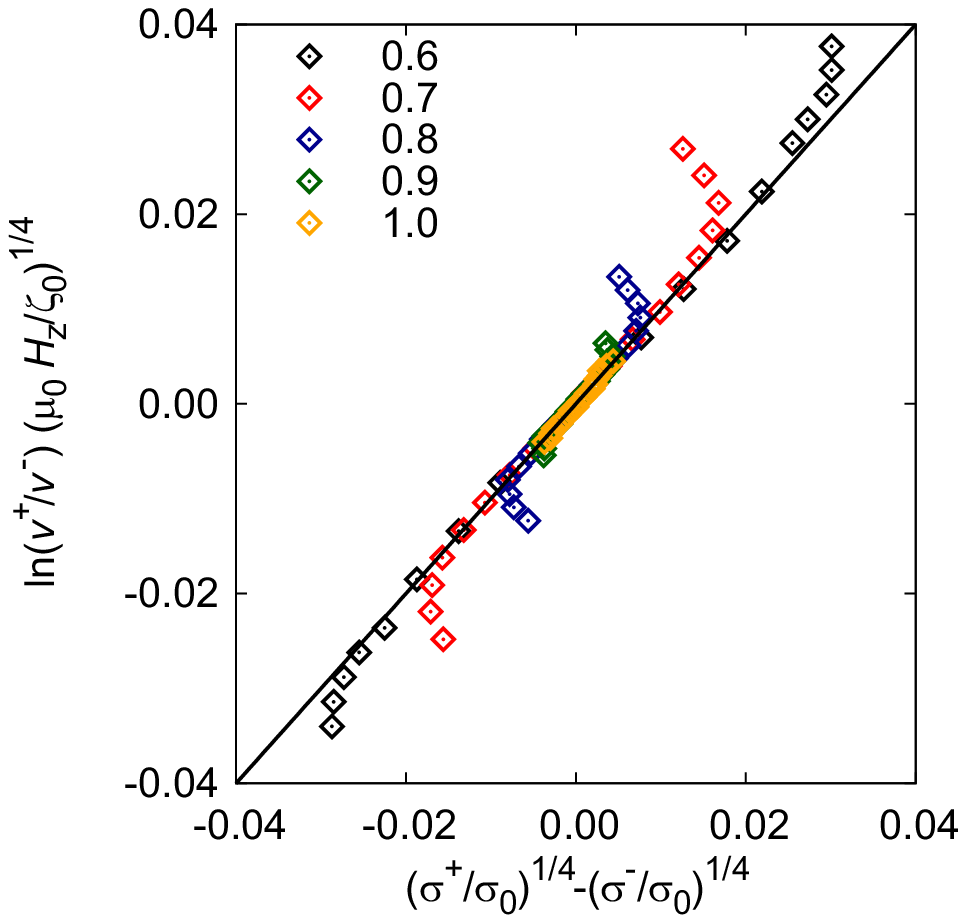}
\caption[Numerical solution for DMI fields]{Dimensionless plot of left and right side of equation~(\ref{eq:num}) relating the experimentally determined velocity ratio to the calculated DMI-induced DW energy reduction. With the correctly adjusted free parameter $H_\text{DM}$ the data points are situated on the plot diagonal. The deviations from the diagonal obtained at larger fields mark the valid data range for the low field analysis.}
\label{Fig:hrabec_num}
\vspace{10pt}
\includegraphics[width=.7\linewidth]{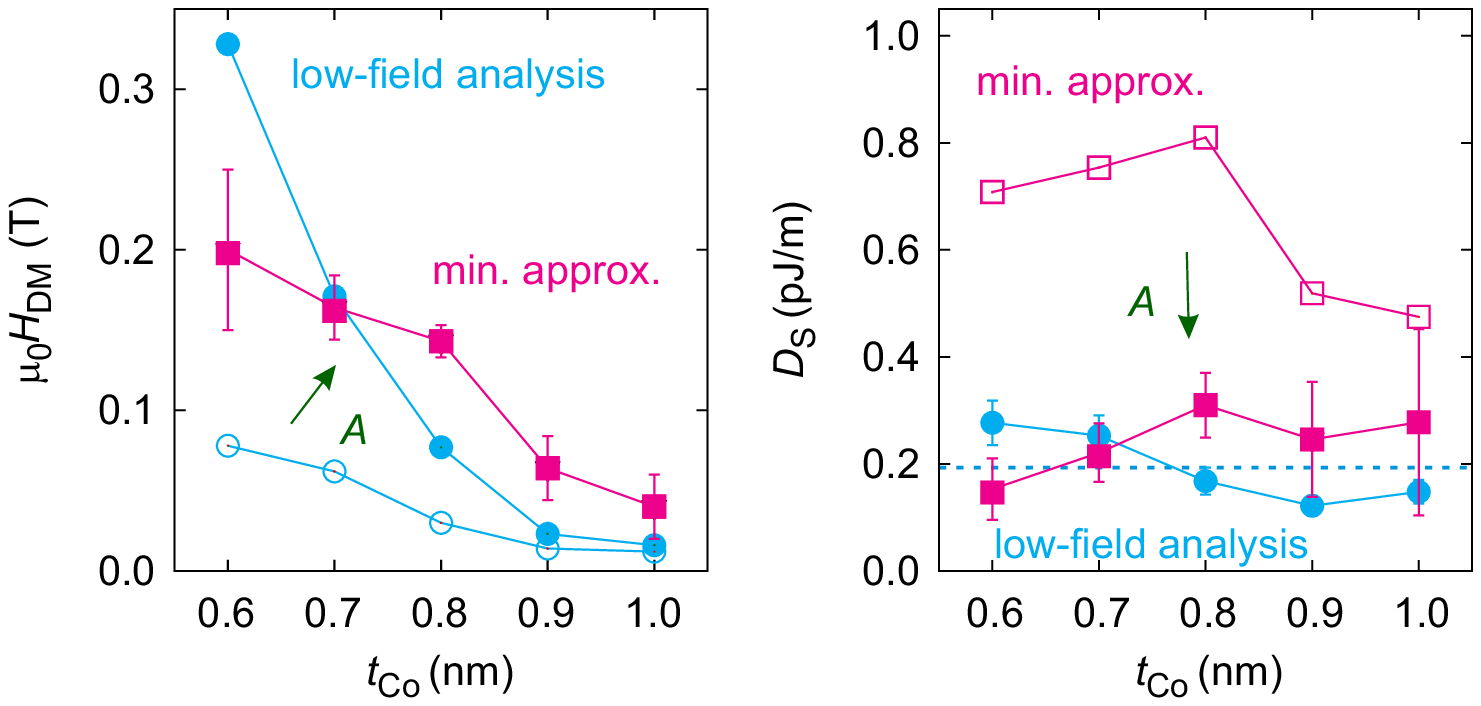}
\caption[Impact of $A$ on the determination of $H_\text{DM}$ and $D_\text{S}$]{\textbf{(a)} DMI fields $H_\text{DM}$ and \textbf{(b)}  thickness-normalized DMI strength $D_\text{S}$ as function of the cobalt thickness $t_\text{Co}$ obtained from  the minimum approximation and the developed low field analysis. The impact on the results is indicated by the green arrow that shifts the values calculated with the bulk-$A$ (unfilled symbols) to those with the correct $A$ (filled symbols) determined from the size of magnetic bubble domains.}
\label{Fig:hrabec_HDM_DS}
\end{figure}

With an additionally field $H_\text{x}$ applied perpendicularly to the easy axis of magnetization, the domain growth exhibits an asymmetric distortion \cite{Kabanov2010} as shown in Fig.~\ref{Fig:hrabec}. This is a consequence of the combined modification of the DW energy by $H_\text{x}$ and the DMI field $H_\text{DM}$ acting in the DW and can be used to deduce the DMI constant from the asymmetry ration of the DW creep velocities $v$ of $\uparrow\downarrow$ (+) and $\downarrow\uparrow$ (-) DWs moving along the axis of $H_\text{x}$ as derived in the methods section of the paper.

The Eq.~(\ref{eq:num}) can be solved numerically for $H_\text{DM}$. 
As shown in Fig.~\ref{Fig:hrabec_num} the dimensionless values of left and right side of Eq.~(\ref{eq:num}) are situated on the diagonal for a correctly adjusted $H_\text{DM}$. 
All points that are deviating from the plot diagonal are not captured by the assumption of the model of sufficiently small $H_\text{x}$ (i.e. the factor $\sqrt{1-(H_\text{x}/H_\text{K})^2}\approx 1$ \cite{Vanatka2015}) and have to be omitted for the analysis.\\

In most of the previous reports \cite{Je2013,Lavrijsen2015,Vanatka2015,Kim2015} it is assumed, that the minimum occurring in $v(H_\text{x})$-curves Fig.~\ref{Fig:hrabec} marks the position where the external field cancels the DMI field. In an approximative approach and only for a limited number of cases $H_\text{DM}$ can then be directly read from a usually broad minimum. This serves as a direct and fast, but only crude measure of $H_\text{DM}$ considering the typically observed data sets, as e.g. shown in Fig.~\ref{Fig:hrabec}.\\
The $v(H_\text{x})$ minimum is usually found in field ranges where the magnetization within the domains is starting to get tilted (i.e. $\sqrt{1-(H_\text{x}/H_\text{K})^2} \ll 1$ \cite{Vanatka2015}). This alters the DW energy by essentially changing the demagnetization factor of the wall and in turn also its mobility. Hence the minimum position cannot be simply identified with $H_\text{DM}$. This becomes increasingly important for thicker films where the anisotropy field is smaller and consequently leads to distortions of the $v(H_\text{x})$-curves (see Fig.~\ref{Fig:hrabec}).\\
Furthermore, the DMI field is a function of all micromagnetic parameters including the saturation magnetization $M_\text{S}$, anisotropy energy $K_\text{eff}$, exchange parameter $A$ and DMI strength $D_\text{S}$. The rescaling to the DMI constant $D_S$ requires a precise knowledge of all these parameters. As $A$ is difficult to determine, usually the bulk value ($A^b$=16~pJ/m for cobalt) is assumed.\\ 
In the following it will be discussed, how the assumption of the bulk exchange parameter $A^b$ can distort the results of the DMI field $H_\text{DM}$ and the DMI strength $D_\text{S}$ for the studied //CrOx/Co($t_\text{Co}$)/Pt samples series. It has been shown that the true exchange parameter in these films is markedly reduced. When using the bulk $A$ in the analysis, the curves of the DMI field $H_\text{DM}$ and the DMI strength $D_\text{S}$ obtained from the minimum approximation and the low field analysis derived in this work are not matching, as shown in Fig.~\ref{Fig:hrabec_HDM_DS}. The experimentally obtained significant reduction of $A$ corrects the deviation between the differently determined $D_\text{S}$ values. The DMI constants calculated from the minimum approximation of $H_\text{DM}$ with the bulk $A$ are clearly overestimated, and are shifted towards the results of the $A$-independent low-field analysis, when the correct exchange parameters are inserted (Fig.~\ref{Fig:hrabec_HDM_DS}b). The DMI fields of the low-field analysis are underestimated when considering the bulk $A$ and are shifted towards the results of the minimum approximation (Fig.~\ref{Fig:hrabec_HDM_DS}a). Despite the $A$-correction the non-matching trends of $H_\text{DM}$ of both methods indicates the approximative nature of the minimum approach. Within the errorbars only one  matching result is obtained for $t_\text{Co}=0.7$~nm. \\
Recently doubts about the commonly applied quantification methods  for the DMI have been raised \cite{Vanatka2015, Soucaille2016}, as the results from the asymmetric magnon dispersion and various quasistatic approaches (Tab.~\ref{Tab:DMIsummary}) yield widely differing values of $D_\text{S}$. The presented results  exemplify one important reason, as the used approaches so far have been unable to determine $D_\text{S}$ and the crucial important exchange constant $A$ independently.\\
The here presented alternative low-field analysis of the field-induced DW velocity asymmetry offers a twofold clarification for the ambiguities of previous results. Firstly one has to be aware of the multi-fold field effects on the DW mobility in different field regions, that can be easily avoided in the low-field approximation. And secondly, the relevance of every single micromagnetic parameter for the  calculation and rescaling of quantities has to be strictly considered.\\
It was further proposed to ascribe the asymmetry of the domain expansion in the picture of a dissipation effect attributed to chiral damping \cite{Jue2016}. According to this idea the DW velocity is inversely proportional to the damping of the system. Depending on the DW chirality the damping is either enhanced or diminished by $\alpha_\text{c}\,l_\text{ex}\,(\mathbf{m}_\text{ip}\cdot \nabla m_\text{z})$, with the exchange length $l_\text{ex}$ and the DW magnetization component $\mathbf{m}_\text{ip}$ perpendicular to the film normal. In contrast to the previously presented approach the attempt frequency $f_0$ in the creep parameter $v_0$, Eq.~(\ref{eq:zeta0}), is modified by the DMI induced chiral damping, while the $\zeta$ is only affected by the hard axis field.  According to Ref.~\cite{Jue2016}, this results in a entirely different qualitative trend of the $v(H_\text{x})$-curves. Contrary to the observation in the here studied samples, no shift along the field axis of this curve is predicted in this model. This precludes the interpretation of the data in terms of the chiral damping. 

\newpage

\section{Generation of circular magnetic objects}
\label{sec:scaling}

\begin{figure}[!p]
\centering
\caption[Magnetic bubble domains in //CrOx/Co/Pt trilayers]{The measured collapse diameter $D_\text{bc}$ of magnetic bubble domains in the experimentally applied field $\mu_0H_\text{ext}$ and the calculated force function $F(d_\text{bc})$ and the stability criterion for the bubble collapse $S_\text{bc}(d_\text{bc})$ from which the bubble collapse field $\mu_0H_\text{bc}$ and the exchange parameter $A$, Eq.~({eq:Aquant}), can be calculated.}
\begin{tabular}{ r|rr|rrr} \hline\hline
\multicolumn{1}{c|}{$t_\text{Co}$}	&	\multicolumn{1}{c}{$D_\text{bc}$}			&	\multicolumn{1}{c|}{$\mu_0H_\text{ext}$}	&\multicolumn{1}{c}{$S_\text{bc}$}	&	\multicolumn{1}{c}{$F_\text{bc}$} &	\multicolumn{1}{c}{$\mu_0H_\text{bc}$} 	\\
\multicolumn{1}{c|}{(nm)}	&		\multicolumn{1}{c}{($\mu$m)}		&	\multicolumn{1}{c|}{(mT)}	&		&		&	\multicolumn{1}{c}{(mT)}	\\ \hline
0.6	&	2.0		$\pm$ 0.4	&	0.06	&	2.864	&	3.182	&	0.14 		\\
0.7	&	1.84	$\pm$ 0.4	&	-		&	2.789 	&			&	-		\\
0.8	&	0.84	$\pm$ 0.20	&	0.29	&	2.496	&	2.815	&	0.52	\\
0.9	&	0.59	$\pm$ 0.20	&	-		&	2.347	&			&	-	\\
1.0	&	0.54	$\pm$ 0.10	&	5.42	&	2.285	&	2.603	&	1.10 \\ \hline\hline
\end{tabular}
\label{Tab:bubble}
\vspace{10pt}
\includegraphics[width=.7\linewidth]{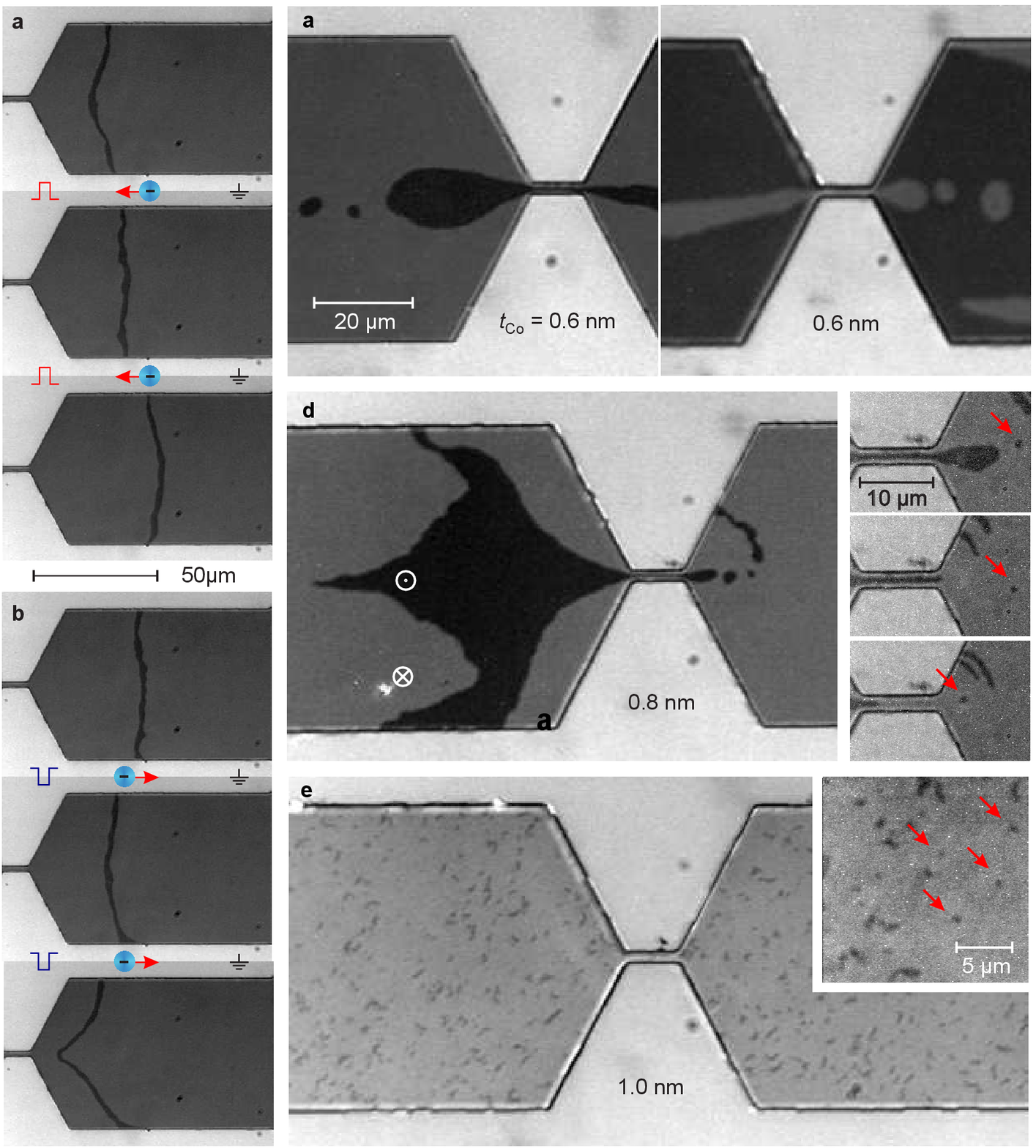}
\caption[Current-induced DW motion and homochiral magnetic circular objects]{Current induced DW motion in a 60~$\mu$m wide //CrOx(5nm)/Co(0.8nm)/Pt(2nm) channel. Both $\uparrow\downarrow$ and $\downarrow\uparrow$ DWs coherently move driven by spin-orbit torques against the electron motion direction with an average velocity of 50~$\mu$m/s at the chosen current density of $j = 5.62\,10^{6}\text{A\,cm}^{-2}$. The images are taken after application of positive \textbf{(a)} and negative \textbf{(b)} pulses of 40~ms length. The velocities vary across the width of the stripe due to the current density profile and are largest close to the constriction of the stripe.
The creation of magnetic circular objects in the scheme of Jiang et al. \cite{Jiang2015} or by a demagnetization routine in //CrOx/Co($t_\text{Co}$)/Pt trilayers with different cobalt layer thicknesses \textbf{(c-e)} indicated by red arrows. Different snapshots tracking the creation process of a the object are shown in panel \textbf{(d)}.}
\label{Fig:bubble}
\end{figure}
Circular chiral magnetic objects may be generated in the scheme of Jiang et al. \cite{Jiang2015} or by applying a demagnetization routine. In the prepared microstripes a coherent current-induced motion of two DWs enclosing a domain is observed upon the application of current pulses to the magnetic stripe (see Fig.~\ref{Fig:bubble}a and b).
The DWs move against the direction of the electron flow, and consequently the DW shift changes sign with the reversal of the current direction. The constriction of the microstripe causes inhomogeneities of the current density which reflects in a distinct DW velocity profile across the microstripe width with a velocity maximum at the center of the stripe (see last image in Fig.~\ref{Fig:bubble}b).\\
The current-induced DW motion is attributed to the action of spin-orbit torques\cite{Emori2013,Ryu2014} on the DWs. It is required that the DWs are homochiral to enable a coherent motion of the object. Additionally, a DW magnetization component parallel to the current direction (i.e. Néel-like type wall) is needed in order to allow for the spin-orbit torques to efficiently induce DW propagation against the direction of electron motion. In the presence of DMI these DW properties are a natural consequence which also result in the formation of winding pairs.\\
Circular homochiral magnetic objects are generated due to the contraction of a domain channel in divergent currents during the expulsion from the constriction area, when additionally applying an easy axis field ($H_z$) to stabilize the created magnetic object (inset of Fig.~\ref{Fig:bubble}d). The resulting objects sometimes are of arbitrary shape and usually shrink to circular objects of a typical diameter. Small alternating external magnetic fields can assist to reach this equilibrium state. Due to large pinning in the samples the objects are rather immobile when applying either magnetic field or current pulses.\\
In the samples with $t_\text{Co}\leq$0.9~nm the strong pinning prevents an efficient current-induced motion. Instead a typical meander pattern of magnetic domains is observed. In these cases the circular objects were observed during the magnetization reversal as remainder of the continuously shrinking meander structures. \\
Considering the scaling of the characteristic diameter of the circular objects with the cobalt layer thickness one can identify them as magnetic bubble domains or skyrmions. The individual scaling behaviors are discussed in the following assuming a constant exchange parameter for simplicity.\\
The diameter of bubbles is related to the ratio $\lambda_\text{c}$ of the DW energy and the shape anisotropy. In the case of the bubble collapse, this ratio $\lambda_\text{c}$ is determined by the stability criterion, which is monotonically increasing with the collapse diameter. The shape anisotropy is a quadratic function of the  saturation magnetization, which decreases almost linearly with decreasing $t_\text{Co}$ in the investigated samples. Therefore $\lambda_\text{c}\propto t_\text{Co}^{-2}$ decreases with increasing $t_\text{Co}$. Hence, cylindrical bubble domains stabilized by the dipolar stray fields are expected to shrink in size with increasing layer thickness, i.e. their size should be inversely proportional to $t_\text{Co}$. \\
On the other hand the skyrmion diameter is approximately proportional to the chiral modulation length $L_\text{D}$) scaling with the ratio of ferromagnetic $A$ to asymmetric $D_\text{eff}$ exchange constant. For decreasing $D_\text{eff}$ with increasing $t_\text{Co}$ an increase of the skyrmion diameter is expected at constant $A$. Chiral skyrmions are therefore expected to become larger in thicker films.\\
The opposing trends allow to identify the character of the observed circular objects. Quantitative estimates then can corroborated the identification, as bubbles are generally much larger than skyrmions under reasonable assumptions about the materials parameters.\\
The observed magnetic objects have diameters of the order of 1~$\mu$m. As shown in Fig.~\ref{Fig:results} their diameters follow the inversely proportional thickness trend violating the skyrmion diameter scaling $\propto (D_\text{S}\,t_\text{Co})^{-1}$. Hence, objects are identified as homochiral bubble domains and the bubble domain theory is employed to determine the remaining unknown magnetic material parameter $A$ from the bubble size (Methods). Tab.~\ref{Tab:bubble} summarizes the experimental and calculated parameters. 
Only if the skyrmion scaling is observed, the exchange parameter can be evaluated in the framework of skyrmion theory. Micromagnetic simulations (Methods) with known $M_\text{S}$, $K_\text{eff}$ and $D_\text{S}$ can be performed to determine the exchange parameter by only varying $A$ until the resulting diameter matches the experimentally observed skyrmion size.\\
In principle a homochiral bubble should be convertible to skyrmion by constriction. Vice versa the expansion of a skyrmion could generate a bubble domain. In the experiments the nucleation of a reverse domain from the collapse site of a bubble domain in opposite field was not observed. This infers that at least one of transformations from homochiral bubbles into skyrmions or vice versa is unlikely to appear in the systems in the chosen experimental condition. Standard Kerr microscopy imaging does not provide the magnification for a visual documentation of this processes given the estimated diameter range for the skyrmions. Together with the hitherto unclarified energetics of the transmutation process between the two objects 
this topic remains unsettled. 

\newpage

\section{DMI discussion}
\label{sec:DMIdisc}

A strongly diverging Pt resistivity in the ultra thin limit of Pt films suggest an increase of Pt/Co interface roughness below 5~nm \cite{Slepicka2008}. It has been established from structural characterization methods that heavy metals are showing an increasingly rough growth for increasing layer thicknesses above 10~nm \cite{Melo2004}. Furthermore, investigations of the DMI strength in systems with ultra-thin Pt cover layers revealed inhomogeneous coverage below 2~nm \cite{Tacchi2017}. The result of these studies suggests the following scenario for the heavy metal film growth. Below the percolation threshold, estimated to be about 2~nm, heavy metals form porous, non-closed layers. This is indicated by almost non-conducting properties \cite{Slepicka2008}, coverage effect in the DMI study of Ref.~\cite{Tacchi2017} as well as the finding of this work that samples with Pt cover layers thinner than 2~nm  are susceptible to oxidation. The continuing improvement of the Pt conductivity up to 6~nm as found in Ref.~\cite{Slepicka2008} suggests a smoothening of the layers before the roughness increases again \cite{Melo2004}.

\begin{figure}[!b]
\centering
\includegraphics[width=.9\linewidth]{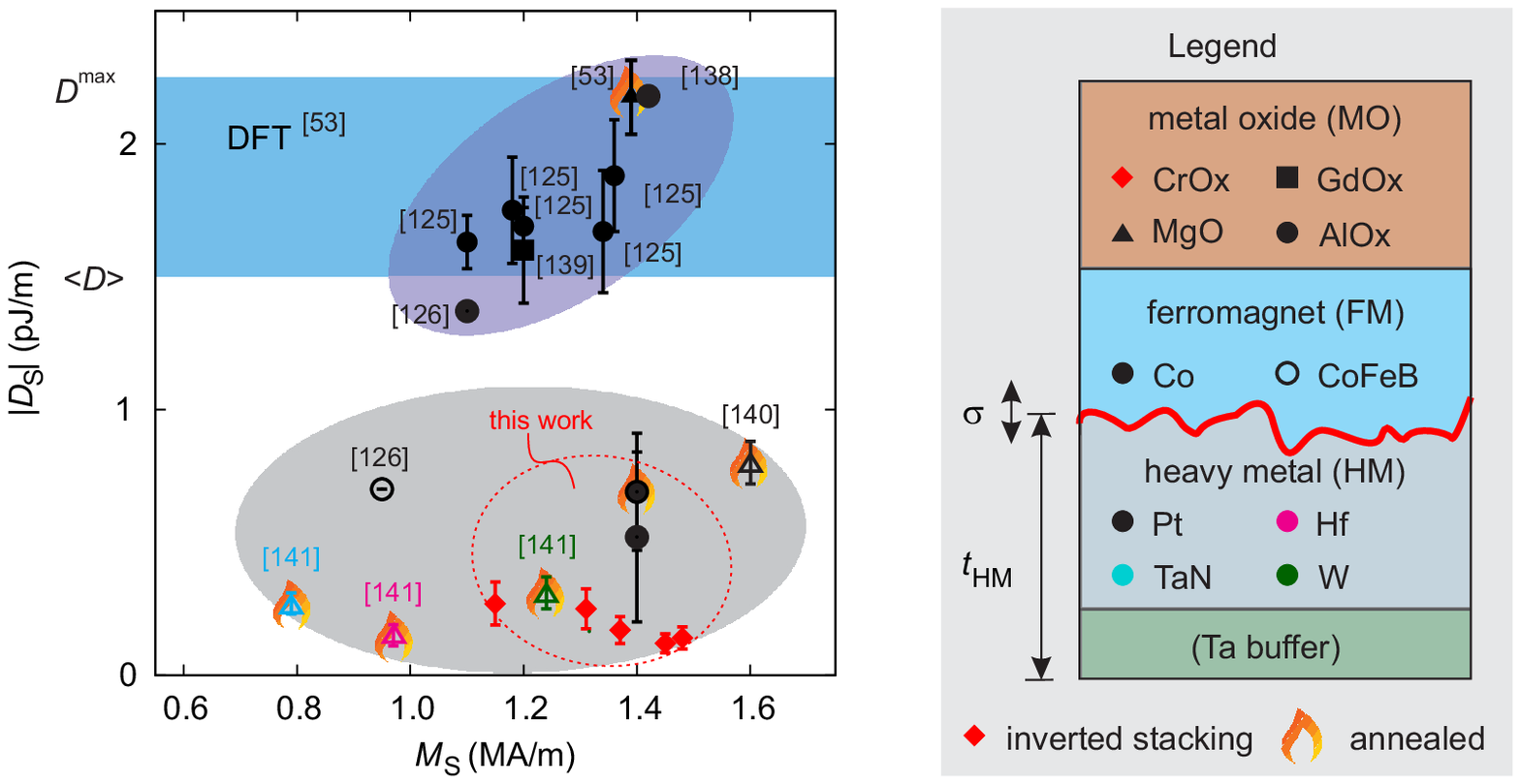}
\caption[Comparison of DMI constants in Pt/Co/metal-oxide trilayers]{Comparison of DMI constants in Pt/Co/metal-oxide trilayers from the literature \cite{Belmeguenai2015, Cho2015, Kim2015a,Vanatka2015, Boulle2016,Di2015a,Soucaille2016} with the results from the studied //CrOx/Co/Pt trilayer and //Pt/Co/AlOx reference samples. Two clusters are forming that gather samples of different microstructural qualities. A first cluster consisting of //Pt/Co/MO trilayers with fairly smooth Co/Pt interfaces at larger DMI values that situates in between the limits of $D^\text{max}$ and $\langle D\rangle$ given by DFT calculations \cite{Boulle2016}. Due to various reasons (see text) structural more complex systems form a second cluster at lower DMI values.}
\label{Fig:DMIcluster}
\end{figure}

Recent ab initio calculations, Fig.~2 of Ref.~\cite{Boulle2016} show a complex distribution of DMI contributions in every atomic layer of Pt/Co and Pt/Co/MgO stacks. Considering the calculation results for Pt/Co interfaces, two limits for the DMI are predicted. A maximum DMI of $D^\text{max}$=2.3~pJ/m is given by taking into account the contribution of the single Co layer adjacent to the Pt layer. The lower limit or averaged DMI of $\langle D\rangle$=1.5~pJ/m is calculated as the sum over all atomic layer contributions for typical Co layer thicknesses (3~ML). It is further concluded \cite{Boulle2016} that the presence of the MgO interface results in an average DMI value exceeding $\langle D\rangle$ of the Pt/Co interface. This is attributed to a contribution of same sign and similar strength arising at the Co/MgO interface promising the ability to enhance the DMI by precise engineering of the FM/MO interface.\\
Figure~\ref{Fig:DMIcluster} contains a summary of data reported so far on HM/FM/MO systems.
The DMI constant $D_\text{S}$ and $M_\text{S}$ are plotted as a phase diagram \cite{Woo2016} to gauge an empirical correlation of the DMI and the details of the sample. A first cluster of comparable Pt/Co/MO samples is formed around an average value of (1.77$\pm$0.27)~pJ/m. Systems with intrinsically different aluminum- \cite{Belmeguenai2015, Cho2015, Kim2015a}, magnesium- \cite{Di2015a,Boulle2016} or gadolinium \cite{Vanatka2015} oxides yield fairly comparable DMI values considering the error bars. 
Therefore, it is concluded, that the Co/Pt-interface is the dominating factor and no clear evidence for an impact of the type of MO on the DMI strength can be found. This result matches with the expectation of weak hybridization of electronic states between Co and MO. In this light the upper cluster lies perfectly within the DFT predicted limits (blue area in Fig.~\ref{Fig:DMIcluster}) suggesting a crucial impact of the Co/Pt interface only. Although a FM/MO contribution might be present, it is not strictly needed to explain the results in the upper cluster. Moreover, it would be required that the DMI contribution arising from the Co/MO interface has the same sign and a similar magnitude independent of the MO (such as Al, Mg or Gd oxides) or its oxidation level.\\
A second cluster is formed at lower DMI values by our results joined by systems comprised of structurally more complex ferromagnets like CoFeB \cite{Di2015a, Cho2015} or different heavy metals (HM) such as Ta, TaN, W and Hf \cite{Torrejon2014, Soucaille2016, Gross2016} in the bottom layer. It has been suggested that the difference in DMI strength for different HMs originate from their respective  electro-negativity \cite{Torrejon2014}. This interpretation might be compromised by the mircostructural details of the interfaces, that are expected to vary with the different heavy metals and were disregarded in Ref.~\cite{Torrejon2014}. In the clarification of this aspect account should be taken of both for the growth conditions of different HMs and the additional complexity of the employed CoFeB ferromagnetic layer. The latter causes undefined changes of the interface properties due to composition variation and/or boron migration to the interfaces \cite{LoConte2015}. At the current state, these systems can not be consulted for the evaluation of the possible DMI mechanism.\\
Nevertheless, assuming a significant contribution of the Co/MO interface the grouping samples with chromium oxide in the lower cluster could then only be explained by an anti-parallel contribution of the CrOx/Co interface partly canceling that of the Co/Pt interface. The obtained experimental results strongly suggest that the DMI in Pt/Co/MO systems dominantly arises from the Pt/Co interfaces, since no empirical evidence for a contribution of the metal-oxide/Co interface to the DMI strength is obtained. The differences of DMI strength between the samples are likely consequences of different interface qualities.

\begin{table}[!t]
\centering
\caption{Data from literature review for Fig. \ref{Fig:DMIcluster}}
\begin{tabular}{l r r r r r r}\hline \hline
stack	&	\multicolumn{1}{c}{$M_s$}		&	\multicolumn{1}{c}{$H_K$}		& \multicolumn{1}{c}{$D_s$} & \multicolumn{1}{c}{$A$} & & 	\\
		&	\multicolumn{1}{c}{(MA/m)}	&	\multicolumn{1}{c}{(MA/m)}	& \multicolumn{1}{c}{(pJ/m)} &   \multicolumn{1}{c}{(pJ/m)} & & 	\\ \hline
//Ta(4)/Pt(4)/Co(t)/AlOx(2)		&	1.42	&	& $	2.18\pm	0	$ 			& 	&	& \cite{Kim2015a} \\
//Ta(3)/Pt(3)/Co(t)/MgO			&	1.40	&	& $	2.17\pm	0.14$ 			& 27.5	&	1.5h, 250${}^\circ$C, vac. & \cite{Boulle2016} \\
//Ta(3)/Pt(3)/Co(0.6)/AlOx(2)		&	1.36	& 0.79	& $	1.88\pm	0.21$ 	& 	&	& \cite{Belmeguenai2015}\\
//Ta(3)/Pt(3)/Co(0.8)/AlOx(2)		&	1.18	& 0.67	& $	1.75\pm	0.20$ 	&	&	& \cite{Belmeguenai2015}\\
//Ta(3)/Pt(3)/Co(0.9)/AlOx(2)		&	1.20	& 0.57	& $	1.69\pm	0.07$ 	& 	&	& \cite{Belmeguenai2015}\\
//Ta(3)/Pt(3)/Co(0.95)/AlOx(2)	&	1.34	& 0.35	& $	1.67\pm	0.23$ 		& 	&	& \cite{Belmeguenai2015}\\
//Ta(3)/Pt(3)/Co(1.2)/AlOx(2)		&	1.10	& 0.08	& $	1.63\pm	0.10$ 	&	& 	& \cite{Belmeguenai2015}\\
//Pt(5)/Co(1)/GdOx(t)				&	1.20	& 0.56	& $	1.60\pm	0.20$ 	& 16 	& 35s O2-plasma & \cite{Vanatka2015} \\
//Pt(4)/Co(t)/AlOx(2)				&	1.10	&	& $	1.37\pm	0	$ 		&	& 	& \cite{Cho2015} \\
//Pt(2)/CoFeB(0.8)/MgO(2)			&	1.60	& 0.19	& $	0.80\pm	0.08$ 	& 	& 1h, 240${}^\circ$C, vac.  & \cite{Di2015a} \\
//Pt(4)/CoFeB(t)/AlOx(2)			&	0.95	&	& $	0.70\pm	0	$ 	&	&	& \cite{Cho2015} \\
//TaN(1)/CoFeB(1)/MgO(2)			&	1.24	&	& $	0.31\pm	0.06$ 	& 	& 1h, 300${}^\circ$C, vac.  & \cite{Soucaille2016} \\
//W(3)/CoFeB(1)/MgO(2)				&	0.79	&	& $	0.27\pm	0.04$ 	& 	& 1h, 300${}^\circ$C, vac.  & \cite{Soucaille2016} \\
//Hf(1)/CoFeB(1)/MgO(2)			&	0.97	&	& $	0.15\pm	0.04$ 	& 	&	1h, 300${}^\circ$C, vac.  & \cite{Soucaille2016} \\ 
\hline\hline
\end{tabular}
\label{Tab:DMIsummary}
\end{table}

\begin{figure}[!t]
\centering
\includegraphics[width=.5\linewidth]{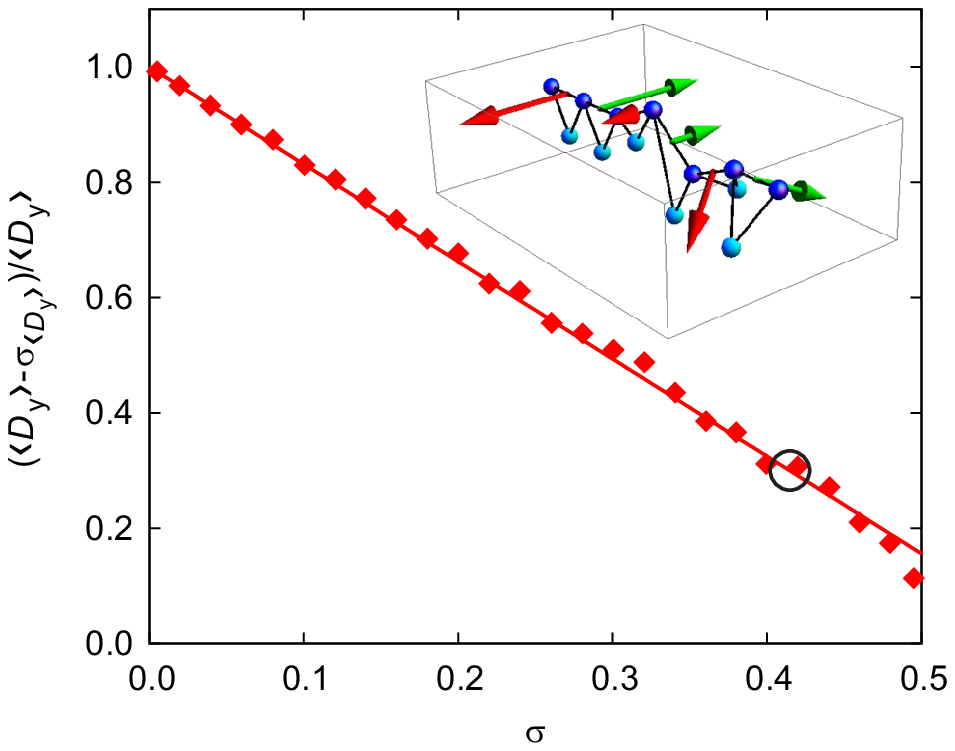}
\caption[Calculated roughness dependence of the DM vector]{DM vector calculated on the basis of 3-sites-model for a linear chain of magnetic atoms (blue/dark gray spheres) on the heavy metal substrate (light gray spheres) as a function of the degree of disorder. The disorder is modeled using Gaussian distribution of interatomic distances with standard deviation $\sigma$. The larger $\sigma$, the stronger disorder. The normalized expression $D_\text{eff}\sim \langle D_y\rangle - \sigma_{\langle D_y\rangle}$ as function of the roughness $sigma$ shows a significant reduction with increasing $\sigma$. Note that this can serve as a qualitative picture, even though the estimated roughness of $\sigma = 0.42$ at 60~\% reduced DMI strength in our samples is reasonable. The inset exemplifies a distribution of the DM vectors.}
\label{Fig:levyfert}
\end{figure}

Sputtered finely polycrystalline thin magnetic films will have various contributions to the chiral DMI, which may compete and eventually average to a vanishing twisting effect, in particular if different surfaces of rotated grains or surface defects like steps cause alternating twisting effects on the magnetization.\\
A random distribution of local DMI terms along the surface with alternating signs and directions of Dzyaloshinskii-vectors $\mathbf{D}$ in the microscopic local DMI between local spins may average to a very weak effective chiral coupling.\\
To provide a cartoon picture of the possible averaging mechanism, the roughness influence on the DMI strength is evaluated by a modified Levy-Fert three-site model \cite{LEVY1981}. To calculate the dependence of the DMI strength on the degree of disorder, the disorder/roughness has been introduced as a Gaussian distribution of the interatomic distances around an ideal lattice constant $a$ with standard deviation $\sigma$ \cite{Kotzler2006}
\begin{equation}\label{Eq:Levy}
    \mathbf{D}_{ij}(\mathbf{R}_{li},\mathbf{R}_{lj})  \propto \frac{\left( \mathbf{R}_{li}\cdot \mathbf{R}_{lj} \right)\left( \mathbf{R}_{li} \times \mathbf{R}_{lj} \right)}{|R_{li}|^3\,|R_{lj}|^3\,R_{ij} }. 
\end{equation}
$\mathbf{R}_{li}$, $\mathbf{R}_{lj}$ are the distance vectors from the impurity $l$ to corresponding FM sites, ${R_{ij}}$ the distance vector between FM sites, while $\mathbf{D}_{ij}$ is corresponding DMI vector. \\
Based on the obtained calculation results the average of the $y$-component of $D$ is found to be constant for all $\sigma$ while the standard deviation of this component linearly increases. It is concluded, that the effective DMI constant $D_\text{eff}$, which is relevant e.g. for the altered DW motion in the experiments, is related to the reduced average of the $D_y$-component by its standard deviation $D_\text{eff}\sim \langle D_y\rangle - \sigma_{\langle D_y\rangle}$. This results in an almost complete destruction of the DMI  at larger disorder as one can see qualitatively from Fig.~\ref{Fig:levyfert}. Even though a constant average $D_y$ should result in the maintenance of chiral order, the large local deviations diminish the effective DMI. 
The inset in Fig.~\ref{Fig:levyfert} shows the distribution of DM vectors in a linear chain of atoms with disorder. Only nearest neighboring magnetic atoms (blue/dark gray spheres) have been considered in the calculations. Each pair has been coupled via a nearest neighboring heavy metal atom (cyan/light gray spheres). For the sake of comparison three left atomic bonds are completely ordered in a hcp-like manner.

\end{document}